\def\simlt{\lower.5ex\hbox{$\; \buildrel < \over \sim \;$}}
\def\simgt{\lower.5ex\hbox{$\; \buildrel > \over \sim \;$}}
\RequirePackage{lineno} 
 \documentclass[preprint2]{emulateapj}

\slugcomment{Accepted by the Astrophysical Journal  June 3, 2011 }
\shorttitle{GJ~436b Atmosphere}
\shortauthors{Line et al.}
\begin{document}
\title{Thermochemical and Photochemical Kinetics in Cooler Hydrogen Dominated Extrasolar Planets: \\
    A Methane-Poor GJ436b?}
\author{Michael R. Line}
\affil{California Institute of Technology, Pasadena, CA 91106}
\author{Gautam Vasisht}
\affil{Jet Propulsion Laboratory, California Institute of Technology, Pasadena, CA 91109}
\author{Pin Chen}
\affil{Jet Propulsion Laboratory, California Institute of Technology, Pasadena, CA 91109}
\author{D. Angerhausen}
\affil{Hamburger Sternwarte, Universit\"at Hamburg, 21029 Hamburg,
    Germany}
\author{Yuk L. Yung}
\affil{California Institute of Technology, Pasadena, CA 91106}
\altaffiltext{1}{Correspondence to be directed to mrl@gps.caltech.edu or gv@s383.jpl.nasa.gov}
\begin{abstract}

We introduce a thermochemical kinetics and photochemical model.
We use high-temperature bidirectional reaction rates for important H,
C, O and N reactions (most importantly for CH$_4$ to CO
interconversion), allowing us to attain thermochemical equilibrium,
deep in an atmosphere, purely kinetically.  This allows the
chemical modeling of an entire atmosphere, from deep-atmosphere
thermochemical equilibrium to the photochemically dominated regime.
We use our model to explore the atmospheric chemistry of cooler
($T_{eff} < 10^3$ K) extrasolar giant planets.  In particular, we
choose to model the nearby hot Neptune GJ436b, the only planet in this
temperature regime for which spectroscopic measurements and estimates
of chemical abundances now exist.  Recent {\it Spitzer} measurements
with retrieval have shown that methane is driven strongly out of
equilibrium and is deeply depleted on the dayside of GJ~436b, whereas
quenched carbon monoxide is abundant.  This is surprising because
GJ~436b is cooler than many of the heavily irradiated hot Jovians and
thermally favorable for CH$_4$, and thus requires an efficient
mechanism for destroying it.  We include realistic estimates of
ultraviolet flux from the parent dM star GJ~436, to bound the direct
photolysis and photosensitized depletion of CH$_4$. While our models
indicate fairly rich disequilibrium conditions are likely in cooler
exoplanets over a range of planetary metallicities, we are unable to
generate the conditions for substantial CH$_4$ destruction.  One
possibility is an anomalous source of abundant H atoms between 0.01-1
bars (which attack CH$_4$), but we cannot as yet identify an efficient
means to produce these hot atoms.

\end{abstract}


\section{Introduction}
Currently, transiting extrasolar planets offer virtually exclusive \footnote[2]{The exceptions to the exclusivity are the few young, self-luminous planets as in the HR8799 system} opportunities
for observing physical and chemical states of exoplanetary
atmospheres.  Over the past four years, retrievals of atmospheric
molecules from multicolor transit photometry (i.e. transit spectra)
have compelled the development of progressively more sophisticated
atmopheric models to interpret the observations and understand
underlying chemical and dynamical processes.  In particular,
atmospheric-chemistry modeling is evolving from strictly
thermo-equilibrium models with stationary chemical species, to coupled
models (Zahnle et al. 2009a,b; Line et al. 2010; Moses et al 2011)
incorporating thermo-kinetics, vertical transport, and photochemistry.
Thus far, such efforts have been devoted to hot-Jupiter planets,
especially HD 209458b and HD 189733b, due to their favorable transit
depths and eclipse brightnesses and, therefore, far greater
availability of observational data.  However, with the recent
retrieval of molecular abundances in the atmosphere of GJ~436b
(Stevenson et al. 2010; Madhusudhan \& Seager 2011), exoplanetary science is
venturing into a new territory: hot-Neptune atmospheric chemistry.
GJ~436b is bound to serve as a prototypical planet anchoring the
theoretical framework for understanding the hot-Neptune class of
exoplanets, much as how HD 209458b and HD 189733b have for
hot Jupiters.  It is also the first planet with observable thermal
emission that transits an M star.  M stars are of particular interest
since they constitute the majority of stars in the solar neighborhood,
and they have close-in habitable zones, which enhances radial-velocity
detectability and transit observability; therefore, M stars present
the best opportunities to discover and characterize rocky, potentially
habitable exoplanets in the near future.  GJ~436b and GJ~1214b provide the only
present test cases for atmospheric chemistry of planets orbiting M
dwarfs.  Therefore, an era of intensive investigations of this planet
is commencing.  This paper presents our application of a
state-of-the-art model seamlessly integrating thermo-kinetics,
vertical transport, and photochemistry to simulate the atmospheric
chemistry of GJ~436b in a similar manner to Visscher et al. (2010) and Moses et al. (2011)
, along with realisitic estimates of UV fluxes for
this planet.

The first transiting hot Neptune discovered (Butler et al. 2004, Gillon et al. 2007),
GJ~436b, revolves around an M dwarf merely 10 pc away from Earth and has received much attention due to its interesting orbital dynamics (Ribas et al. 2008, Mardling 2008, Batygin et al. 2009), interior properties (Nettelmann et al. 2010, Kramm et al. 2011), and atmospheric properties  (Stevenson et al. 2010, Lewis et al. 2010, Madhusudhan \& Seager 2011, Shabram et al. 2011).  The
slightly eccentric orbit (eccentricity = 0.16) has a mean orbital
radius of 0.0287 AU (Torres et al. 2008), and the planet probably has
a pseudo-synchronous rotation (Deming et al. 2007).  The planet's mass
is 23 $M_{\oplus}$, and its density of 1.7 g/cm$^3$ resembles that of
the ice-giant Neptune (1.63 g/cm$^3$).  Analyses of its mass-radius
relationship and transit depth indicates a layer of H/He dominated
atmosphere is clearly required (Figueira et al. 2009; Nettelmann et
al. 2010; Rogers \& Seager 2010).  The host star has an effective
temperature of $\sim$~3400 K and an estimated age of 3 -- 9 Gyr
(Torres et al. 2008).  Assuming zero albedo and global thermal
re-distribution, the planet's effective temperature is 650 K.  Of the
confirmed transiting exoplanets (Wright et al. 2011), GJ~436b is one
of the least irradiated and has one of the coolest atmospheres.
Therefore, this planet represents a significant departure from
hot Jupiters in terms of size, thermal environment, and UV flux.

Although GJ 436b was discovered in 2004 (Butler, by radial velocity),
it was not until 2010 that a retrieval of explicit molecular
abundances in its atmosphere was reported (Stevenson et al. 2010),
where six channels of secondary-eclipse photometry data ranging from
3.6 to 24 $\mu$m were analyzed by generating $\sim$ 10$^6$ simulated
spectra using varying combinations of molecular compositions and
temperature profiles to find the best fit to observations.  A more
recent paper (Madhusudhan \& Seager 2011) provides further details and
updated results of a re-analysis of the same dataset using the same
general retrieval method.  In short, 10$^6$ combinations of ten
physio-chemical free parameters, each spanning a large range of
values, were used to generate synthetic dayside-emission spectra.  In
each of the 10$^6$ scenarios, six of the ten parameters were used to
define the temperature-pressure (T-P) profile, whereas the other four
parameters specified vertically uniform abundances of four
molecules: H$_2$O, CO, CH$_4$, and CO$_2$.  Additionally, the 1-D
atmospheric model restricted the ratio of emergent flux output to
incident stellar flux input on the day side to within the range
between zero and unity.  Given six data points and ten free paramters,
the retrieval problem was mathematically underdetermined.
Nonetheless, sampling a million points in parameter-phase space
allowed the authors to examine the joint probability contours, as
defined by the goodness-of-fit (chi-square) function, projected on
multiple-parameter spaces.  Furthermore, by placing
physical-plausibility constraints (in consideration of believable
departures from thermo-equilibrium chemistry) on the molecular
abundances, the authors were able to confine the physical space to a
fairly narrow, ``best-fit,'' range for chi-square $\le$ 3.  Depending
on the wavelength, the photospheric altitude varies from 9 bar to 0.2
bar levels.  The main conclusions are as follows: 1) temperature
inversion is ruled out (i.e. no stratosphere); 2) 6 ppm (parts per
million) is the absolute upper limit for CH$_4$ abundance; 3) 300 ppm
is the absolute upper limit for H$_2$O abundance; 4) CO$_2$ and CO
abundances are anti-correlated; 5) taking
physical-plausibility into consideration, the best-fit spectrum
represents $X_{\rm H_{2}O}$ = 100 ppm, $X_{\rm CH_{4}}$ = 1 ppm, $X_{\rm CO}$ =
7000 ppm, and $X_{\rm CO_{2}}$ = 6 ppm, where $X_{\rm i}$ is the number density
of molecule $i$ divided by that of H$_2$.  Also, note that even in the
best-fit scenario, $X_{CO_2}$ can range anywhere from 1 -- 100 ppm.
The Stevenson et al. (2010) and the Madhusudhan \& Seager (2011)
efforts are the most comprehensive studies of atmospheric composition
on GJ~436b thus far.

From a theoretical point of view, the preceding abundance limits and
values pose a very interesting challenge due to their drastic
departures from thermo-equilibrium predictions, which indicate the
following rough-order-of-magnitude values: $X_{\rm H_{2}O}$ = 1000
(3$\times$10$^4$) ppm, $X_{\rm CH_{4}}$ = 1000 (10$^4$) ppm, $X_{\rm CO}$ = 60
(10$^4$) ppm, and $X_{\rm CO_{2}}$ = 0.1 (1000) ppm for 1x (50x) solar
metallicities at $\sim1$bar.  In either metallicity scenario, water
and methane remain abundant ($\ge$ 1000 ppm), whereas the retrieval
shows water being relatively depleted and methane being drastically
depleted.  Moreover, the thermo-equilibrium abundances of carbon
monoxide and carbon dioxide are positively correlated (either both low
in the 1$\times$ case or both high in the 50$\times$ case), in contrast with the
retrieval's anti-correlation.  In particular, the retrieved partioning
of carbon overwhelmingly in oxidized species amidst a
hydrogen-dominated (reducing), temperate atmosphere is very
surprising.  For instance, at 1-bar pressure and solar metallicity,
CH$_4$ is the thermodynamically dominant carbon-bearing molecule for
temperatures less than 1100 K (Lodders \& Fegley 2002).  The common
practices of simply adjusting metallicity and/or the C/O ratio cannot
simultaneously reconcile these discrepancies.  Therefore, one must
investigate disequilibrium mechanisms.

Madhusudhan \& Seager (2011) posited that high metallicity combined
with vertical mixing can explain the disequilibrium abundance of
carbon oxides.  Basically, enhanced metallicity ($\sim$ 10$\times$ solar) can
provide the requisite abundance of CO$_2$.  Since
equilibrium CO abundance drops sharply with respect to temperature
(Lodders \& Fegley 2002) the retrieved uniformly high abundance of CO
requires eddy mixing to populate upper, cooler, atmospheric layers.
However, vertical eddy mixing alone cannot explain the large depletion of
CH$_4$ due to its innately high thermochemical abundance in the deep atmosphere.  
Therefore,  Madhusudhan \& Seager (2011) invoked photochemistry as the potential
culprit, based on Zahnle et al.'s (2009a,b) studies of photochemistry
on hot Jupiters.  In such a scheme, photosensitized sulfur chemistry
produces atomic H, which then destroys CH$_4$ to form higher
hydrocarbons.  However, the Zahnle et al. (2009a,b) model uses
solar-type stellar irradiance and an isothermal atmsophere
(i.e. constant temperature versus altitude).  As such, neither the
photochemical driver nor the thermal environment is tailored for our
planet in question.  More severely, Moses et al. (2011) pointed out
that a typo in a key rate coefficient in the Zahnle et al. (2009a,b)
model caused the apparent conversion of methane into higher
hydrocarbons at pressures larger than 1 mbar.  Generally speaking,
at pressures larger than 1 mbar in a hydrogen-abundant atmosphere,
hydrogenation of unsaturated hydrocarbons and reaction intermediates
efficiently recycle species back to methane, preventing its large-
scale destruction.  Moses et al. (2011) also discussed the inadequacies
of isothermal atmospheric models due to their suppression of
transport-induced quenching.  Hence, the observed CH$_4$ depletion
still awaits adequate explanation.  The low abundance of H$_2$O also
has not been addressed.

In addition to secondary eclipse observations, primary transit observations of GJ 436b exist as well 
(Pont et al. 2009, Ballard et al. 2010, Beaulieu et al. 2011, Knutson et al 2011),
and various groups have analyzed them to retrieve molecular abundances in the planet's terminator regions 
(Beaulieu et al 2011, Knutson et al 2011).  
In contrast to the secondary-eclipse retrieval, 
Beaulieu et al. (2011) were able to fit a compendium of their 
and Ballard et al.'s tansit observations between 0.5 and 9 $\mu$m with 500 ppm CH$_4$ in a H$_2$ atmosphere, 
and finding no clear evidence for CO or CO$_2$.  
Moreover, Beaulieu et al. presented that a methane-rich atmosphere, with temperature inversion, can be 
consistent with the said secondary-eclipse data as well (but see Shabram et al. 2011).  More recently, Knutson et al. 
acquired Spitzer transit photometry at 3.6, 4.5, and 8.0 $\mu$m during 11 visits.  The multiple-visit data 
showed high transit-depth variability, which the authors attribute to potential stellar activity in the dM host.  They did not find any 
compelling evidence for methane, and data excluding ones believed to be most affected by stellar activity 
appear to place an upper limit of 10 ppm for methane mixing ratio.  The best-fit spectrum to this select data set
assumes 1000 ppm H$_2$O, 1000 ppm CO, 1 ppm CH$_4$, with CO$_2$ abundance poorly constrained, roughly 
in agreement with Madhusudhan et al.  Therefore, primary-transit data is currently inconclusive due to 
different interpretations by different groups.

Our primary goal is to advance the fundamental understanding of
processes impacting the chemical state of GJ~436b by developing a 1-D
atmospheric model that integrates all of the aforementioned
equilibrium and disequilibrium processes.  An important aspect of our
model is the seamless integration of thermochemistry, kinetics,
vertical mixing, and photochemistry in a manner that directly follows from Visscher et al. (2010), and contemporaneously with
Moses et al.  (2011), obviating the conventional quench-level
estimation (Prinn \& Barshay 1977).

The quench-level approach assumes that the deep atmosphere is in
thermochemical equilibrium because high temperatures provide
sufficient kinetic energy to overcome reaction barriers in either
direction.  However, as vertical tranport lifts a gas parcel to
cooler, higher altitudes, chemistry becomes rate limited rather than
thermodynamically determined.  There comes a point in altitude where the
kinetic conversion time scale becomes slower than the transport time
scale, and the rate-limiting reaction for a molecule of interest is
not allowed time to reach completion.  At altitudes above this point,
the molecule's concentration is frozen/quenched (therefore, the term
``quench level").  In effect, the quench-level approach partitions the
atmosphere into two parts: below the quench level, thermochemical
equilibrium determines chemical abundances; above the quench level,
molecular abundances are uniform versus altitude, with values equal to
the equilibrium value at the appropriate quench level for each
species.  Although this approach has a long record of success
(e.g., Prinn \& Barshay 1977; Smith 1998; Griffith \& Yelle 1999; Saumon et al. 2003; 2006; 2007; Hubeny \& Burrows 2007; 
Cooper \& Showman 2006), it does have some limiting assumptions and
caveats that require great judiciousness.  Specifically, one needs to
determine the appropriate rate-limiting, interconversion reaction
for each set of coupled species of interest
(e.g., interconversion between CH$_4$ \& CO).  The correct reaction
choice is not always readily apparent (see e.g., Visscher et al. 2010)
and the appropriate length scale for deriving the mixing time
scale from the vertical eddy diffusion coefficient ($K_{zz}$) is still
under some debate.  Furthermore, since a basic assumption is that
temperature decreases with altitude, atmospheric temperature
inversions can complicate matters.

Therefore, we implemented a fully reversible kinetic model in the
following manner.  Every measured forward reaction rate in our list is
reversed using the equilibrium constant and the principle of
microscopic reversibility.  Given enough pathways, both forward and
backwards, a given set of chemical species will reach
thermochemical equilibrium, kinetically.  This provides a seamless
transition from the thermochemical equilibrium regime to the
disequilibrium-dominated regimes.  We can investigate the
disequilibrium effects on atmospheric composition in a much more
holistic, systematic manner, compared to heuristically identifying
plausible disequilibrium processes.

In the remainder of this manuscript we describe the disequilibrium
processes that may be occurring in GJ436b's atmosphere.  In $\S$2 we
describe thermochemical and chemical kinetics models as well as our
estimate for the stellar UV flux.  In $\S$3 we show the modeling
results as well as a description of the important reaction schemes
governing the abundances of various species.  Finally in $\S$4 we
discuss the relevant implications and conclude.

\section{Description of Models}

We use joint thermochemistry and ``1-D chemical-kinetics with
photochemistry'' models to study the atmosphere's departure
from thermal equilibrium. External inputs to our models are the metals fraction (denoted further on by $\zeta$), 
the pressure and temperature (T-P) profile, the eddy diffusion coefficient profile, and the incident stellar flux; 
note that we fix the T-P profile and the chemistry is decoupled from it, 
i.e., there is no self-consistent, radiative-convective adjustment of 
temperature structure when the chemistry is evolved towards steady state.
We initialize the 1-D atmospheres using the NASA Chemical Equilibrium
with Applications (CEA) model (Gordon \& McBride 1996). 
Given the initial elemental abundances of H,
He, C, O, N, and S in an atmospheric layer, along with the layer's
pressure and temperature, CEA uses a Gibbs free-energy minimization
and mass balance routine to calculate the equilibrium species
abundances. 

Whereas chemical equilibrium concentrations are useful for initializing the
atmosphere, they do not provide the correct chemical state above
pressure levels of $\simeq 10$ bars (Prinn \& Barshay 1977; Griffith \&
Yelle 1999; Cooper \& Showman 2006; Line et al. 2010; Moses et
al. 2011). We simply supply the equilibirum mixing-ratios
as boundary conditions in the deep atmosphere for the kinetics calculations, and
thereafter evolve the chemical state over multiple timesteps until a
steady state is reached.

The computations are carried out with the Caltech/JPL photochemical
and kinetics model, KINETICS (a fully implicit, finite difference
code), which solves the coupled continuity equations for each involved
species, and includes transport via molecular and eddy diffusion
(Allen et al. 1981; Yung et al. 1984; Gladstone et al. 1996; Moses et
al. 2005). We use the H, C, and O chemical reaction list originally
described in Liang et al.(2003; 2004) and references therein updated to high temperatures, recently
augmented with a set of N reactions.  We have not
included the chemistry of sulfur in any great detail, because much of
its kinetics is poorly constrained (see e.g., Moses et al. 1996).  However we
do consider a small, but well measured, set of H$_2$S reactions. This helps
us appraise if and how the introduction of S affects the abundances of 
the main molecular reservoirs of H, C, N, O such as CH$_4$.

We use high temperature rate coefficients for reactions from Line et
al. (2010). All reactions are bidirectional, and we reverse them by
calculating the back-reaction rates using thermodynamic data (see
Table S1).  With appropriate reaction pathways and proper rates for the  back-reactions, the 
models can converge to chemical equilibrium purely kinetically in the deep
planetary atmosphere where reaction timescales are short
compared to transport timescales, and photochemical reactions are
unimportant. As mentioned earlier, this removes the cumbersome requirement of having to
choose a lower boundary for individual species through {\it ad hoc}
quench-level arguments (Prinn \& Barshay 1977; Smith et al. 1998).

We solve for 51 hydrogen, carbon, oxygen and nitrogen bearing species
including H, He, H$_2$, C, CH, $^1$CH$_2$, $^3$CH$_2$, CH$_3$, CH$_4$,
C$_2$, C$_2$H, C$_2$H$_2$, C$_2$H$_3$, C$_2$H$_4$, C$_2$H$_5$,
C$_2$H$_6$, O, O(1D), O$_{2}$, OH, H$_2$O, CO, CO2, HCO, H$_2$CO,
CH$_2$OH, CH$_3$O, CH$_3$OH, HCCO, H$_2$CCO, CH$_3$CO, CH$_3$CHO,
C$_2$H$_4$OH, N, N$_2$, NH, NH$_2$, NH$_3$, N$_2$H, N$_2$H$_2$,
N$_2$H$_3$, N$_2$H$_4$, NO, HNO, NCO, HCN, CN, CH$_3$NH$_2$,
CH$_2$NH$_2$, CH$_2$NH, H$_2$CN, with a total of $\sim 700$ reactions,
55 of which are photolysis reactions. The chemical pathway for
reducing CO to CH$_4$, described recently for Jupiter's deep
atmosphere (Visscher et al. 2010), is included in our reaction list,
along with the reverse pathways for CH$_4$ to CO
oxidation. Photolysis absorption cross sections
are from Moses et al. (2005) and the thermodynamic data (i.e. the
compilation of entropies and enthalpies) used to reverse the kinetic
rate coefficients are from JANAF and CEA thermobuild databases;
e.g., CEA uses data from Chase et al. (1998) and Gurvich et al. (1989)
(see Zehe et al. 2002).

\subsection {Model Parameters}

We model a large pressure and altitude range, 10$^3$ to 10$^{-11}$ bars ($\sim$5000 km or $\sim$0.2 R$_p$ from the 1 bar level),
 so as to capture the three major atmospheric regimes and the 
transitions between them. These three dominant portions of the atmosphere
are -- the thermal equilibrium regime in the deep hot atmosphere, the
eddy transport dominated regime at intermediate pressures, and the
photochemical regime at low pressures. A total of 190 pressure levels,
uniform in logarithmic space, are used between the abovementioned
levels, giving a resolution of about 14 levels per decade of pressure. 
Altitudes above the homopause remain relatively cool in our models, and
we disregard the possibility of a hot thermosphere despite the
models extend up to exophere levels at 10$^{-11}$ bars; this simplification has
little or no bearing on the state of the atmosphere below the
homopause ($P \sim1 \mu$bar).  We adopt the $\zeta=1$ T-P profile 
from Lewis et al. (2010) (see Figure 1),  noting its similarity to the
T-P profile retrieved in Madhusudhan \& Seager (2011) and Stevenson et al. (2010).  
Whereas GJ~436 itself is slightly subsolar in abundances (Bean et al. 2006),  we allow for a span of planetary 
metallicities, covering the cases
$\zeta = 0.1, 1, 50$, and allowing for the possibility that the planet is either enriched or 
depleted; we used solar abundances from the standard text of Yung \& DeMore
(1999)\footnote[2]{Yung \& DeMore (1999) tabulate the abundances of Anders \& Ebihara (1982). These
values predate the more recent downward revision of elements C, O etc. in
the Solar photosphere (reviewed in Asplund et al. 2009). Our  C/H, O/H, N/H and S/H ratios are a  
factor 1.66, 1.52, 1.35 and 1.43 higher than those recommended in Asplund et al. (2009). On this
revised scale we are modeling a planet with $\zeta \simeq 0.16, 1.6, 80$. This was brought to our 
attention by the anonymous referee.}.
For non-solar atmospheres we tune the fractions of C, N, O, and S relative to H but not relative
to each-other (e.g., C/O, N/O, S/O, are always fixed).

\begin{figure}
  \centering
    \includegraphics[width=0.5\textwidth]{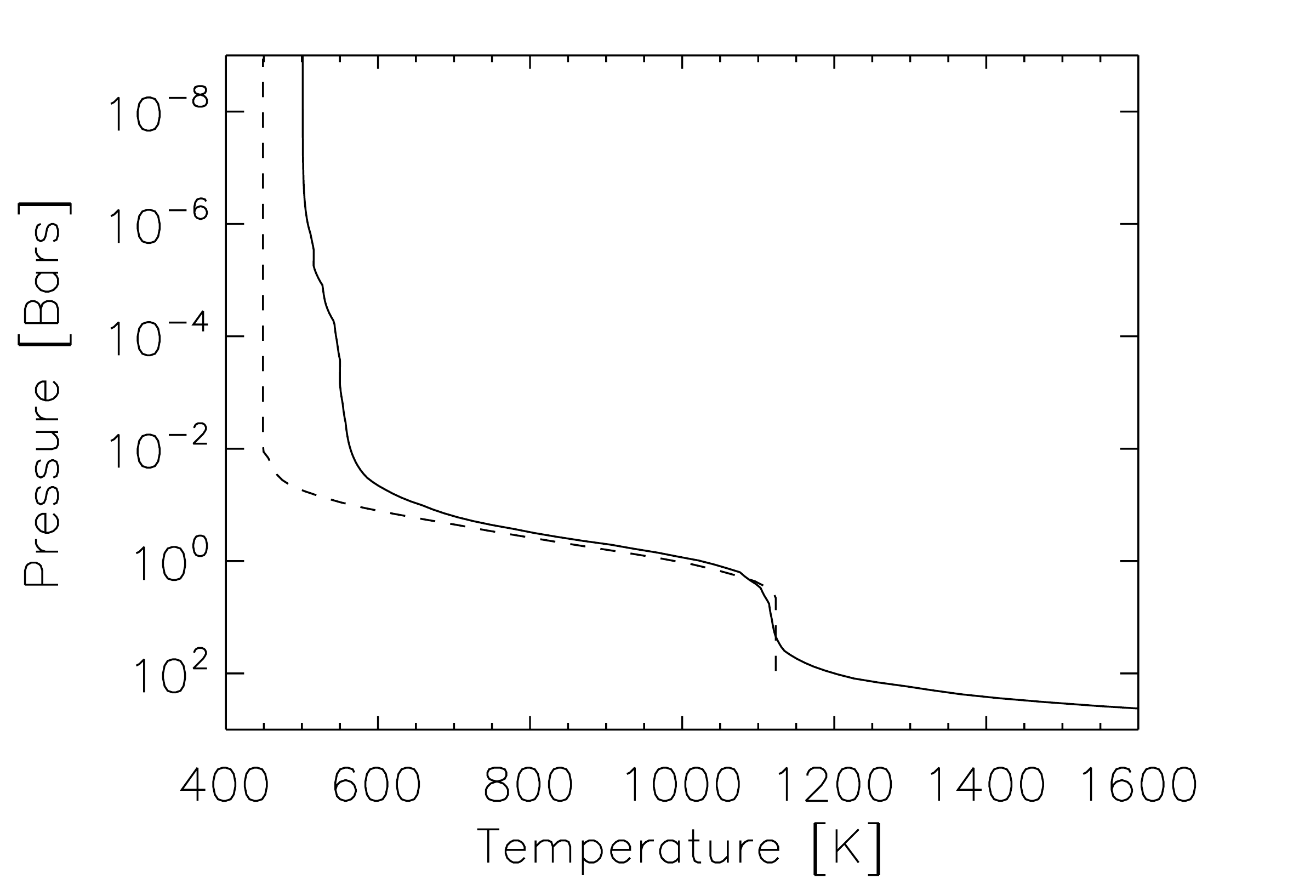}
     \caption{Estimated temperature profiles for GJ~436b.  
     The dashed profile is the disk averaged dayside profile retrieved by Stevenson et al. (2010). 
     The solid curve is the $\zeta=1$ profile from the global circulation model of Lewis et al. (2010).  We use the latter T-P profile 
     for our chemical models.}
\end{figure}

The eddy diffusion strength (parameterized by a coeffcient, $K_{zz}$)
determines the pressure level at which a species is chemically
quenched.  At the quench level for chemical $X$, the timescale for
vertical transport ($\tau_{trans}$) equals the chemical loss timescale
($\tau_{chem,X}$).  Above that level, which includes the visible
portion of the atmosphere, the mixing virtually ``freezes'' the
concentration of that species.
Below the quench level, $\tau_{chem,X} \ll \tau_{trans}$, and
thermochemical balance is achieved.  Line et
al. (2010) and Moses et al. (2011) have used piecewise estimates of the 
eddy diffusion profiles, $K_{zz}(P)$. The recipe has been to estimate
$K_{zz}$ in the deep adiabatic troposphere ($\sim 10^3$ bars) using mixing length theories 
(e.g., Flasar \& Gierasch 1977) and stitch this to 
global circulation model (GCM) derived profiles obtained by multiplying the (horizontally averaged) 
GCM vertical winds of Showman et al. (2009) by the local scale height.  
Lewis et al. (2010) apply this procedure to their GJ~436b
circulation model, and estimate that $K_{zz}$ increases from $ \sim
10^{8}$ at depth (100 bars) to $10^{11}$ cm$^2$ s$^{-1}$ at lower
pressures (1 mbar). 

Such procedures have gnawing uncertainties
-- for example, the appropriate eddy mixing length may only be a fraction
of the scale height, or the vertical wind strengths could well be
overestimated.  Smith (1998) has demonstrated theoretically that using
an eddy length scale equal to the scale height is inappropriate, and may
lead to gross over-estimates of the length scale ($L$) and the timescale 
($\tau_{trans} = L^2/K_{zz}$).  Herein, we simplify matters 
by choosing a constant $K_{zz}(P) =
10^{8}$ cm$^2$ s$^{-1}$ profile;  this value is
similar to that for the deep atmosphere in the Lewis et al. GCM. This
simplification has a couple of redeeming features.
First, this gives quench levels similar
to those that would be derived had we used a GCM-inspired $K_{zz}$
profile. Second, whereas a low $K_{zz}$ may underestimate the mixing strength at
higher altitudes, it has the effect more lethargic replenishment of methane
and other photodissociated species from the lower atmosphere (it bolsters the
photochemical timescale, relative to $\tau_{trans}$).

\subsection {The Ultraviolet Emission from GJ~436}

dM stars such as GJ~436 show very little photospheric emission in the
near to far ultraviolet (UV).  Nevertheless, non-radiative energetic
processes can transport energy to power a hot outer atmosphere, and
this energy is partially dissipated in the form of cooling,
chromospheric UV emission. Because the UV emission levels depend on
many factors, {\it ab initio} estimates of it are difficult.  We use
GALEX and ROSAT derived estimates for GJ~436 and combine these with a
$T_{eff} \simeq 3400$ K continuum from the stellar photosphere.  This
combined emission is used to drive photochemical reactions in GJ~436b.

In the planetary atmosphere both H$_2$ and He are weak absorbers
relative to other molecular species, but are enormously more abundant.
Helium ceases to absorb longwards of $500$ \AA, and H$_2$ longwards of
$1000$ \AA.  Methane, a carbon reservoir and the molecule of
particular interest herein, has a large absorption cross-section
shortwards of 1600 \AA.  Whereas methane (and water) is largely
shielded by H$_2$ and He from very shortwave radiation, it is
photodissociated by radiation between 1000-1600 \AA, and is therefore
susceptible to possible intense H I Ly $\alpha$ ($\lambda = 1216$ \AA)
from the M star host. Longwards of $\lambda = 1600$ \AA, direct
photolysis of methane dwindles due to a combination of the falling
cross-section and weak stellar flux.  Hydrogen sulfide
photodissociates at much longer wavelengths, $\lambda < 2600$ \AA, and
if present in substantial quantities, is poorly shielded by other
reservoir molecules H$_2$, CH$_4$, H$_2$O, etc.  H$_2$S photolysis and
the resultant hot atomic hydrogen may be influential if
$\lambda \simeq 2600$ \AA~photons can penetrate deep into the
planetary atmosphere (more in \S 3.3.5).

GJ~436 is detected in a GALEX survey exposure in the near UV channel
with flux $f_{nuv} = 21.0\pm 3.7$ $\mu$Jy (near-UV channel,
$\bar \lambda = 2267$ \AA, $\Delta \lambda_{FWHM} = 616$ \AA). It is
undetected in the GALEX far UV band, with a $3\sigma$ upper limit of
$f_{fuv} \le 24$ $\mu$Jy (far-UV channel, $\bar \lambda = 1516$ \AA,
$\Delta \lambda_{FWHM} = 270$ \AA). These can be converted to incident
UV photon fluxes at the mean orbital separation of GJ~436b. The near
UV detection implies a flux of $9\times 10^{10}$ photons cm$^{-2}$
s$^{-1}$ \AA$^{-1}$ $\lambda = 1960-2580$ \AA~at the planetary
substellar point.  This dosage at GJ~436b is about 0.2 PELs
(present-Earth-levels); mean Solar photon flux at Earth is $4.7\times
10^{11}$ photons cm$^{-2}$ s$^{-1}$ \AA$^{-1}$ between
2000-2500 \AA~(Yung \& DeMore 1999).  The $3\sigma$ flux upper bound
(GALEX far-UV channel) is $\le 1.3 \times 10^{11}$ photons cm$^{-2}$
s$^{-1}$ \AA$^{-1}$ $\lambda = 1450-1650 \AA$; this is just a factor of
two higher than present-Earth-levels in an equivalent passband.

H Ly $\alpha$ emission can be powerful in the upper chromospheres of
cool stars.  Because it is strongly absorbed in the interstellar
medium, direct line strength estimates are difficult.  We make
an indirect determination based on empirical correlations with soft
X-ray fluxes.  Soft X-ray emission from GJ~436 has been observed in the
Rosat All Sky Survey (H\"unsch et al. 1999), with $f_x \simeq
5.4\times 10^{-14}$ erg cm$^{-2}$ s$^{-1}$ (0.1-2.4 keV; Rosat PSPC),
implying a fractional X-ray luminosity of $L_x/L_{bol} \sim 8\times
10^{-6}$; this fraction is a factor $\sim 100$ lower than that
observed from the most active dM stars and is consistent with GJ436b's
estimated advanced age, $6\pm3$ Gyr.  More recent XMM-Newton EPIC
measurements (Sanz-Forcada et al. 2010) give a factor of 8 lower
$L_x$, which may well be due to X-ray activity. Herein, we adopt the
ROSAT flux because larger X-ray fluxes imply
proportionally larger Ly $\alpha$ fluxes.

To estimate the Ly $\alpha$ output, we use an an empirical correlation
of the X-ray and Ly $\alpha$ emission of stars, derived from stellar
samples that include several late type stars (e.g., Landsman \& Simon
1993 and Woods et al. 2004; in these papers, measurements of Ly
$\alpha$ lines were made from {\it International Ultraviolet Explorer}
and {\it Hubble Space Telescope} spectra, after applying a model based
correction of ISM absorption).  Inverting the Woods et al. (2004)
empirical power law, $\log F_x \simeq 2.2\log F_{Ly\alpha} - 7.76$, we
determine a photon flux of $f_{Ly\alpha} \sim 1.5\times 10^{14}$ photons cm$^{-2}$
s$^{-1}$ at GJ~436b\footnote[3]{Very recently, Ehrenreich et al. (2011) estimate a Ly $\alpha$ flux 
using HST-STIS observations of GJ436.  Their estimated line flux is a factor 
$1.5\times$ smaller than the estimate based on L$_x$ used
herein}. The Solar H Ly $\alpha$ flux at Earth is $\simeq 10^{12}$ photons cm$^{-2}$
s$^{-1}$, a factor 100 lower.
The reliability of X-ray derived Lyman $\alpha$ line flux may be assessed
by comparing $F_{Ly\alpha}$ with the GJ~436b's H $\alpha$ line flux.  H $\alpha$ 
observed in GJ436 in absorption, with an equivalent width of
0.32 \AA~ (Palomar-Michigan State Nearby Star Spectroscopic Survey; Gizis,
Reid \& Hawley 2002), implies a line
flux of $F_{H\alpha} \simeq 2\times 10^5$ erg cm$^{-2}$ s$^{-1}$, and
a line strength ratio of H Ly
$\alpha$ to H $\alpha$ of 2.2. For dM stars, where H Ly $\alpha$ is seen
in emission and for which the intrinsic Ly $\alpha$ line strengths
have been measured, this line strength ratio varies between
3-5, with some stars having ratios as low as 2 and others as high as 8 (Doyle
et al. 1997).

\section{Chemical Model Results}
\subsection{Thermochemical Equilibrium}

\begin{figure}
  \centering
    \includegraphics[width=0.5\textwidth]{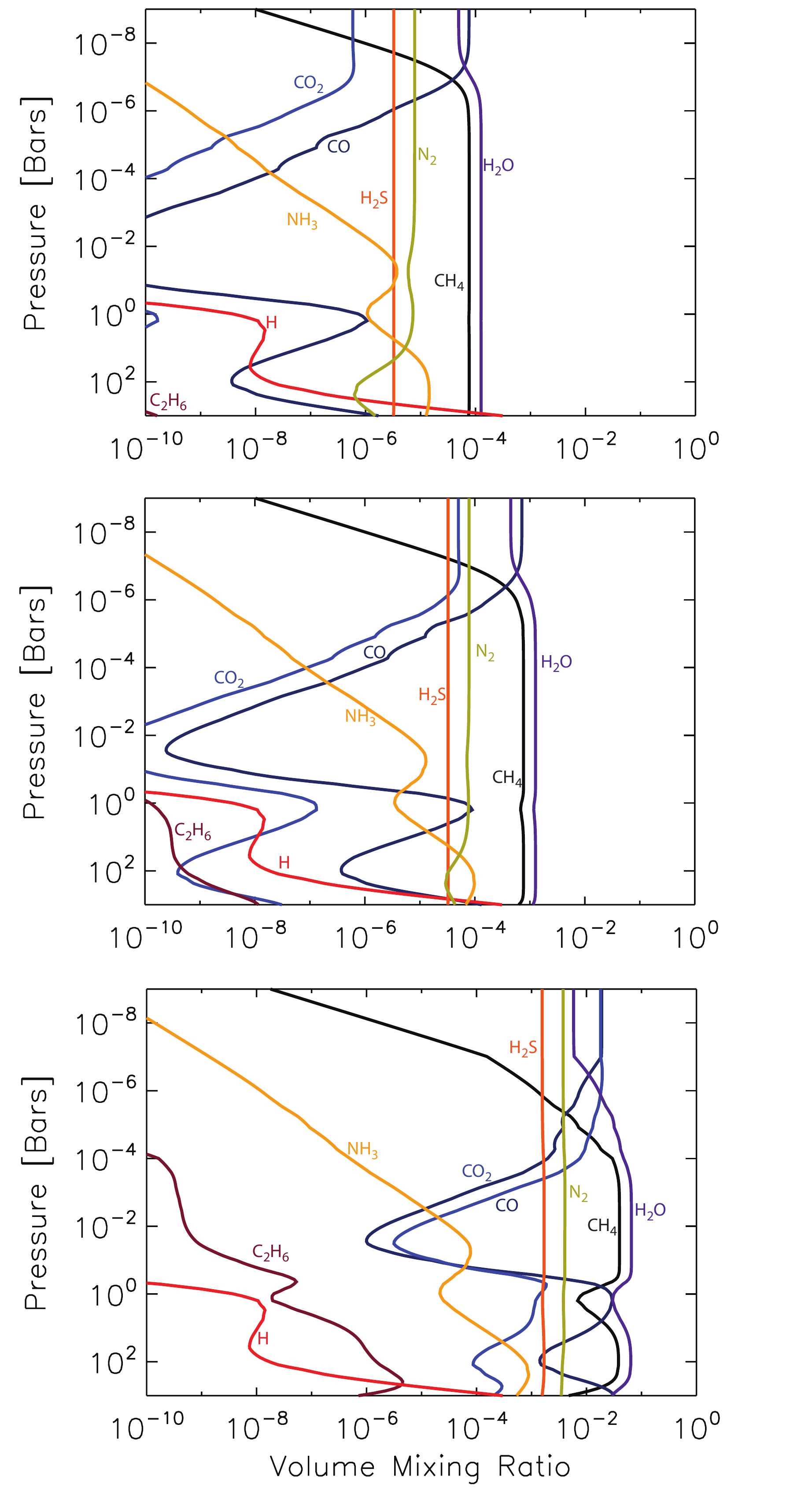}
     \caption{Thermochemical equilibrium vertical distributions for abundant H, C, O, N, and S species 
     assuming the temperature profile in Figure 2.  Three metallicity cases are shown ($\zeta =0.1$, $1$, and $50$, from top to bottom).  
     The thermochemical equilibrium mixing ratios are derived using the CEA Gibbs free energy minimization code for each atmospheric T-P level.}
\end{figure}

Equilibrium vertical mixing ratios for the three metallicity cases are
shown in Figure 2: these are sub-solar $\zeta = 0.1$, solar $\zeta = 1$
and super-solar $\zeta = 50$ heavy elemental abundances. Because GJ~436b
is significantly cooler than HD~189733b and HD~209458b, CH$_4$ is
the thermochemically favored carbon carrier; higher effective temperatures drive
equilibrium towards CO in the two hot Jupiters.  The thermochemical
abundances of CH$_{4}$, CO and H$_{2}$O along the T-P profile are
readily understood through the net reaction
\begin{equation}
{\rm CO + 3H_{2}} \rightleftharpoons {\rm CH_{4} + H_{2}O}
\end{equation}
along with the Law of Mass Action:
\begin{equation}
	\frac{X_{CH_{4}}X_{H_{2}O}}{X_{CO}X^{3}_{H_{2}}}\frac{1}{P^{2}}=K_{eq}(T)
\end{equation}
derived by minimizing the Gibbs free energy of net reaction in (1),
with the mixing ratio $X_{i}$ of species $i$, with ambient pressure $P$, 
and a temperature dependent equilibrium constant $K_{eq}(T)$; the 
$T$ dependence is governed by the van 't Hoff equation ($\Delta G = -RT \log K_{eq}$,
with $\Delta G$ as the standard Gibbs free energy change).  At a
given pressure $P$,  $K_{eq}(T)$ behaves in a manner that
rising $T$ drives the equilibrium towards CO.
At a fixed $T$, increasing/decreasing
pressures favor higher CH$_4$/CO concentrations.  These relationships
are exemplified in the $\zeta = 1$ equilibrium profiles shown in Figure 2 (middle panel). As
$P$ and $T$ decrease along the adiabat between $1000 - 100$ bars, the 
equilibrium constant dominates over the adverse $P^2$ dependence, resulting in
a drop in the CO fraction.  In the isothermal region between $10 - 1$ bars, decreasing
pressure now favors the production of CO. Between 1
bar and $\approx 10^{-2}$ bars, the CO fraction falls because of the rapid decrease in temperature with altitude.
At levels above the $\sim 10^{-2}$ level the temperature structure is nearly isothermal, and the decreasing
pressure favors higher CO fractions.  Similarly, NH$_3$ is
the favored N carrier deep in the atmosphere, but is less favored at
lower atmospheric pressures.  Sulfur can be predominant as
H$_2$S, HS, or S depending on pressure and temperature, but for
conditions prevalent in GJ~436b, gas phase H$_2$S is the dominant sulfur
reservoir and its concentration is unaffected by the
temperature structure.  Heavier
hydrocarbons, such as ethane (C$_2$H$_6$), are relatively scarce any pressure or
temperature (but more common at the highest metallicities).

Enriching the atmosphere to $\zeta = 50$ increases the mixing ratios of
the reservoir species in proportion, however the shapes of the vertical profiles
are much the same as for solar metallicities.  Similarly, decreasing the
metallicity of the atmosphere to $\zeta = 0.1$ lowers the 
mixing ratios of the heavy gases, by a factor $\sim \zeta$ for CH$_4$ and
$\zeta^2$ for CO etc. The shapes of
vertical distributions are nonetheless preserved, and relatively insensitive to 
$\zeta$.

For all three metallicity cases considered, the chemical equilibrium abundances of CH$_4$ and H$_2$O stay relatively high
-- there is always enough hydrogen present to build these molecules.
One can imagine an extreme situation where H is highly depleted,
but such an atmosphere would be incompatible with the observed planetary radius.   
Conversely, the planet could be impoverished in metals 
to greatly subsolar levels $\zeta \ll 0.1$, although unreasonably low metallicities 
($\le 1\times10^{-5}\times$ solar) would be required to 
deplete CH$_4$ and other common molecules to levels below 1 ppm.  
These simple cases serve to show that, based solely on 
chemical thermodynamics, CH$_4$ has to be
relatively abundant in GJ~436b and other $T_{eff} = 500 - 1000$ K H-rich planets.

\subsection{Vertical Mixing \& Chemical Quenching}
Vertical turbulent mixing has been  invoked to explain the anomalously large observed
abundance of CO in Jupiter (Prinn \& Barshay 1977) and brown dwarfs such
as GL~229b (Griffith \& Yelle 1999). Diffusive tropospheric mixing, in
combination with detailed CO chemistry, has recently been used to 
infer the water inventory in the deep Jovian atmosphere (Visscher et al. 2010).
Cooper \& Showman (2006) parameterized the quench chemistry of 
CH$_4$ in order to study its horizontal and vertical transport in their GCM of 
HD~189733b. The recent paper by Moses et al. (2011) discusses in 
detail the quench chemistry of H,C,N,O molecular species in the relatively
hot atmospheres of HD~189733b and HD~209458b. 

In our kinetics models we set thermochemical abundances as boundary
conditions;  these equilibrium abundance
boundary conditions also define the metallicity of the system.  We affix the
$10^3$ bar mixing ratios of the large carbon, oxygen and nitrogen
reservoirs, CH$_{4}$, H$_{2}$O, CO, N$_{2}$, and NH$_{3}$, at their
thermochemically derived values (here we are excluding sulfur), and
set all other species to obey a zero flux condition at the lower boundary.
The exact location of this lower boundary is
unimportant, provided it is at depths much greater than the
quench level ($\ge 100$ bars), and conditions (the high densities and
temperatures) favor thermochemical 
equilibrium concentrations for practically all species.  
The nominal case has a solar abundance atmosphere ($\zeta = 1$), vertical mixing 
with strength $K_{zz} = 1\times 10^8$ cm$^{2}$ s$^{-1}$, and no photochemistry.  In Figure 3 we
compare an atmosphere with vertical mixing to one purely in equilibrium. Below
10s of bars, the mixing ratios converge, satisfying the condition that
equilibrium concentrations have been reached kinetically. 
Now consider the
abundances of quenched CO. At pressure levels deeper than 10s of bars, the eddy mixing time, $\tau_{trans}$, must be
longer than the chemical loss timescale. As a check for internal consistency,
we estimate
\begin{equation}
\tau_{trans} = {L^2 \over K_{zz}} \simeq  8\times 10^5~{\rm s}
\end{equation}
where $L$ is a fraction $f$ of the scale height $H$, $L = fH$ (Smith et al. 1998).
We estimate $f=0.3$ for both quenched CO and N$_2$.
To estimate $\tau_{chem}$ for CO, we need to identify the rate-limiting
reaction in CO and CH$_4$ interconversion. 
\begin{center}	
$$\rm{H + CO + M \rightarrow HCO + M} 	\eqno{R605}$$
$$\rm{H_2 + HCO \rightarrow H_2CO + H} \eqno{R234}$$
$$\rm{H + H_2CO + M \rightarrow CH_3O + M} \eqno{R611}$$
$$\rm{H_2 + CH_3O \rightarrow CH_3OH + H} \eqno{R351}$$
$$\rm{H + CH_3OH \rightarrow CH_{3} + H_{2}O }\eqno{R295}$$
$$\rm{H_{2} + CH_{3} \rightarrow CH_{4} + H }  \eqno{ R61}$$
\rule{2.5in}{1pt}
$$\rm{Net: 3H_{2} + CO \rightarrow CH_{4} + H_{2}O} \eqno{I}$$
\end{center}
This set of reactions is identical to the ones identified for CO quenching in
Jupiter (Yung et al. 1988; Visscher et al. 2010).   The
rate-limiting reaction is R351, the inverse of a hydrogen abstraction
from methanol.
The chemical loss timescale for CO is, 
\begin{equation}
\tau_{chem,\rm{CO}}=\frac{[\rm{CO]}}{k_{351}[\rm{H_{2}}][\rm{CH_{3}O}]}
\end{equation}
where $[X]$ denotes the concentration X, and
$k_{351}=2.10\times10^{-25} T^{4.0} e^{-2470/T}$ cm$^3$ mol$^{-1}$ s$^{-1}$ (Jodkowski et
al. 1999) the rate coefficient for R351.  Figure 4 shows that equality of these
two timescales, $\tau_{chem,CO} \approx \tau_{trans}$,
gives a CO quench-level of $\sim$30 bars, which  furthermore
agrees well with quench-level depicted by the CO mixing ratio profiles in Figure 3.

  \begin{figure*}
  \centering
    \includegraphics[width=0.9\textwidth]{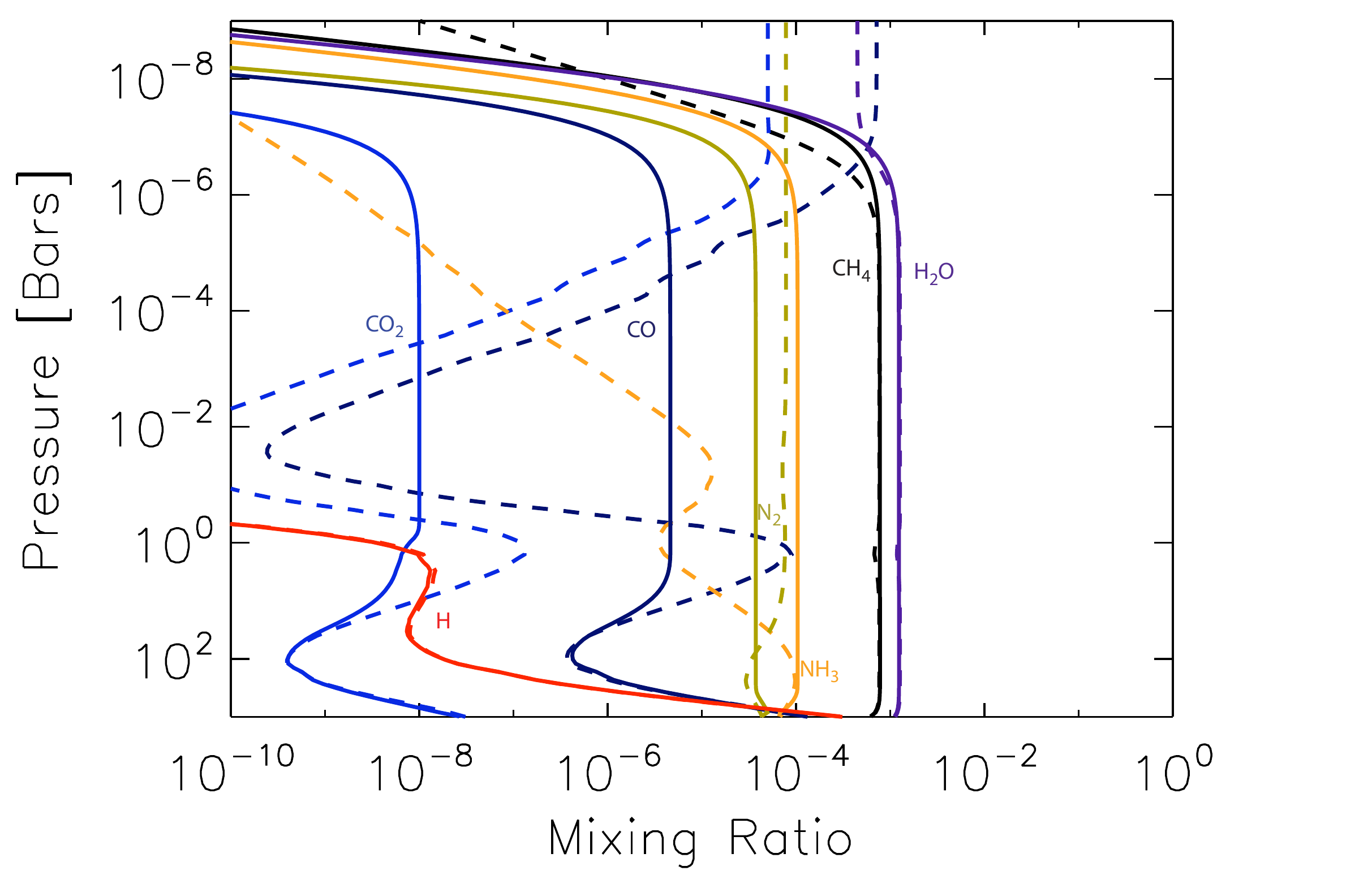}
     \caption{Effects of vertical mixing on the  distributions of  H$_2$O, CH$_4$, NH$_3$, N$_2$, CO, CO$_2$, and H.   The dashed curves are the thermochemical equilibrium profiles for the $\zeta=1$ case from Figure 2 (middle panel).  The solid curves are the vertical profiles derived kinetically with only eddy ($K_{zz}=1\times10^{8}$ cm$^{2}$ s$^{-1}$) and molecular diffusion (no photochemistry) using the 1000 bar $\zeta=1$ mixing ratios as the lower boundary condition.  Note that kinetically derived profiles begin to match the thermochemical equilibrium profiles at levels below a few 10's of bars.  The rapid fall off of the solid curves near 1 $\mu$bar is due to the sedimentation of the heavier molecules because of molecular diffusion. }
\end{figure*}

In an analogous manner the N$_2$ quench-level may be calculated by
identifying the rate-limiting step in the series of reactions that
convert nitrogen to ammonia, and vice versa. 
These reactions are
\begin{center}
$$\rm {H + N_2 + M \rightarrow N_2H + M}  \eqno{R629}$$ 
$$\rm{H_2 + N_2H \rightarrow N_2H_2 + H } \eqno{R478}$$
$$\rm{H_2 + N_2H_2 \rightarrow NH_2 + NH_2} \eqno{R450}$$
$$\rm{ 2(H_2 + NH_2 \rightarrow NH_3 + H)}  \eqno{R453}$$
\rule{2.5in}{1pt}
$$\rm{Net: 4H_2 + N_2 \rightarrow 2NH_3 + 2H }\eqno{II}$$
\end{center}
In this $\rm{N_2 \rightleftharpoons NH_3}$ sequence R450 is the rate-limiting step, involving
the $\rm{N}$ abstraction from diazene, giving a timescale
\begin{equation}
\tau_{chem, \rm{N_{2}}}=\frac{[\rm{N_{2}}]}{k_{450}[\rm{H_{2}}][\rm{N_{2}H_{2}}]}
\end{equation}
with reaction rate $k_{450}=2.06\times10^{-07} T^{-0.93} e^{-20614/T}$, obtained from
that of its reverse reaction (Stothard et al. 1995). 
Calculating $\tau_{chem,\rm{N_2}}$ above gives a 
N$_{2}$ quench-level of $\sim$300 bars (see Figure 4), in agreement with 
the vertical profiles in Figure 3. The abovementioned quench-levels for CO
and N$_2$ are for the adopted eddy diffusion coefficient, $K_{zz} = 10^8$ 
cm$^2$ s$^{-1}$.  Increasing $K_{zz}$ to a very large value, $10^{11}$ cm$^{2}$ s$^{-1}$, 
shortens the transport times considerably and  
increases the quench pressures of CO and N$_{2}$ to $\sim 150$ bars and $\sim
620$ bars, respectively. The effects of varying the quench level 
may be seen in Figure 2 -- the atmospheric concentrations of the 
reservoir gases, CH$_4$, H$_2$O and NH$_3$, and quenched N$_2$, are relatively insensitive to
the location of quench pressure. However, varying the
quench-level affects the concentration of 
CO and CO$_2$ by orders-of-magnitude.  

Vertical dredging of gases leaves a reasonably altered composition
in the 1-0.001 bar region, the range of pressure levels wherein the 
infrared photosphere is located (e.g., Knutson et al. 2009; Swain et al. 2009). 
For example CO is up to
a factor $10^4$ more abundant than it would otherwise be. The deep
quenching of N bearing gases causes NH$_3$ to be surprisingly abundant, 
dominating over the thermochemically favored N$_2$. In contrast, the 
largest C and O reservoirs and optically the most 
active gases, CH$_4$ and H$_2$O, are largely unaffected.

 \begin{figure}
  \centering
    \includegraphics[width=0.5\textwidth]{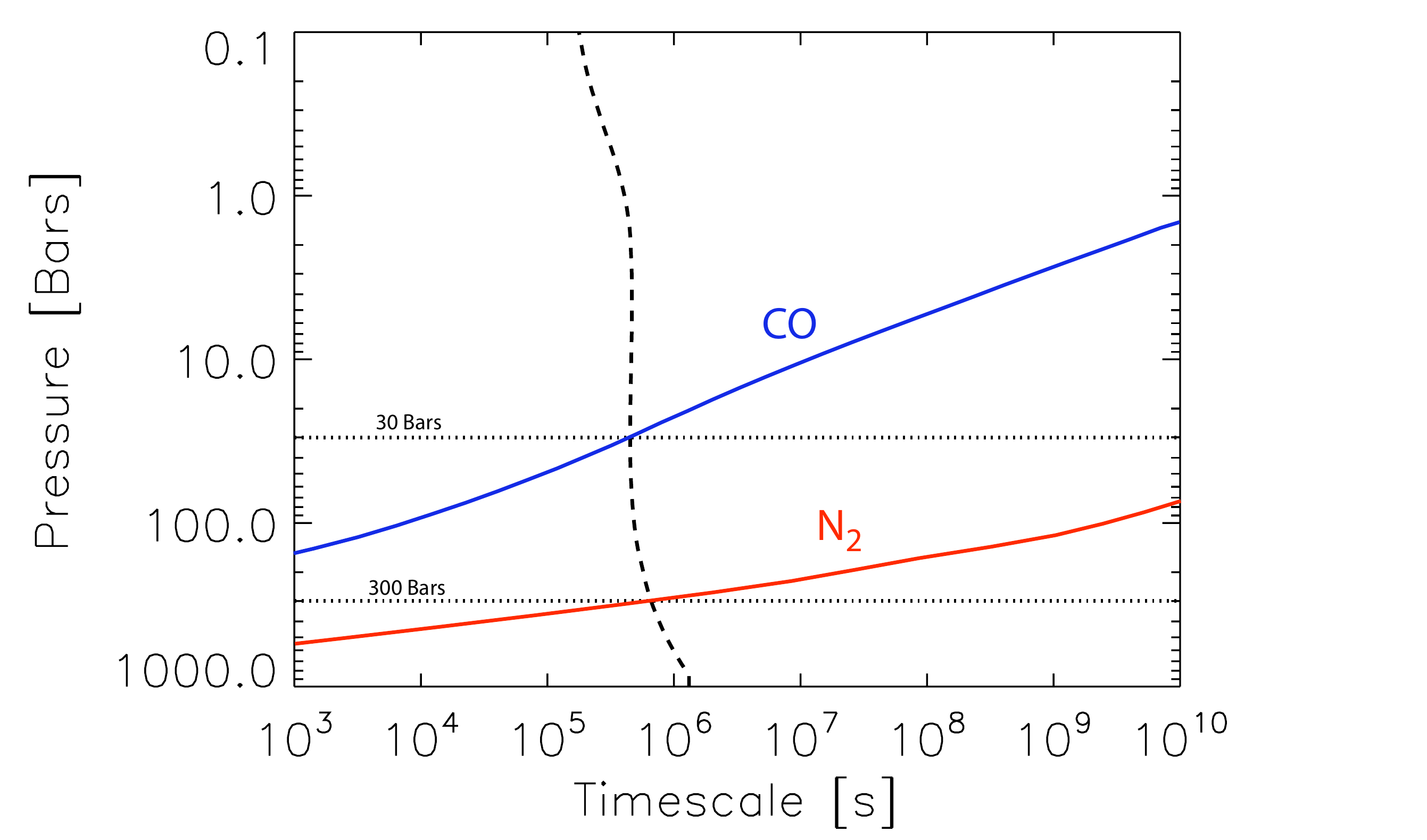}
     \caption{ The blue curve is the CO chemical loss timescale calculated from Equation 4.  The red curve is the N$_2$ chemical loss timescale calculated from Equation 5.  The dashed curve is the vertical mixing timescale from Equation 3  using a length scale of $\sim0.3$H estimated from the Smith et al. (1998) procedure..  The intersection of the vertical mixing timescale and the chemical loss timescale is the quench level for the given species as indicated by the horizontal dotted lines.}
\end{figure}

\subsection{Photochemical Effects}

Photochemistry can significantly alter  atmospheric composition in the
upper portions.  The combination of the ultraviolet flux and molecular
absorption cross sections gives the photolysis rates for all the species
considered here.
The altitude of peak production/loss 
(in units of cm$^{-3}$ s$^{-1}$), set by the balance between the exponential fall-off of
atmospheric density and the inward stellar UV attenuation, occurs near 1 $\mu$bar (this is the well known Chapman function, see Yung \& DeMore 1999 pg. 45).  
Primarily, photolysis breaks apart stable molecules into radicals, which  
can then react to alter the composition of the upper atmosphere.  See Figures 5, 6 and 7 for the
photochemically derived mixing ratios.  Table 1 compares the column
mixing ratios from our models to the observations over the 7 bar to
0.1 bar range probed by the observations.  
Figure 8 illustrates how photochemistry alters the upper atmosphere. 
The resultant mixing ratio profiles are compared with those obtained via thermochemical 
equilibrium (Figure 2), and by vertical mixing (Figure 3).
\begin{figure}
  \centering
    \includegraphics[width=0.5\textwidth]{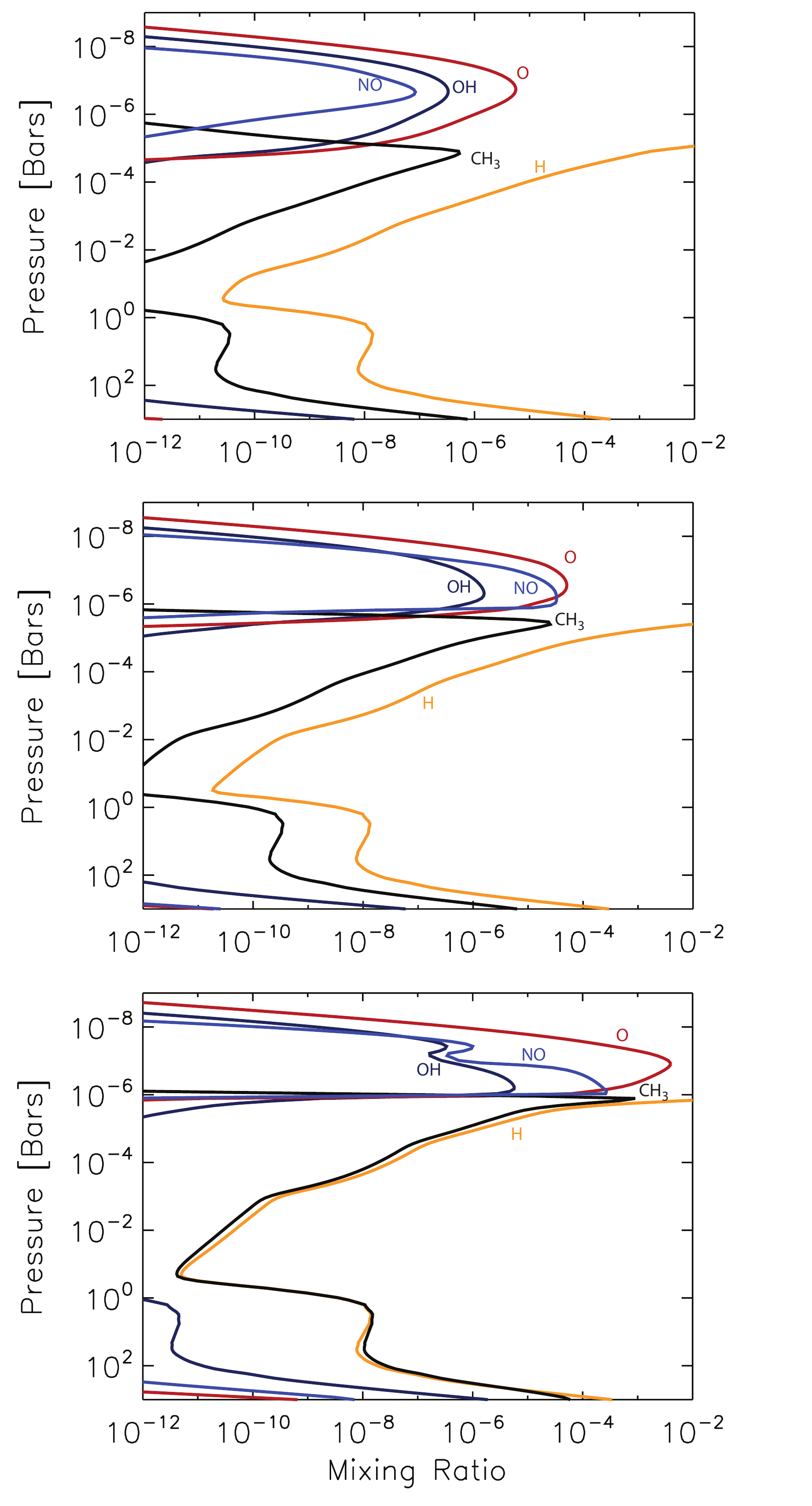}
     \caption{Mixing ratios for important radicals  (OH, NO, O, H, and CH$_3$) that drive the photochemistry for three metallicities ($\zeta=0.1$(top), $\zeta=1$ (middle), and $\zeta=50$ (bottom)). Note how the CH$_3$ profile very nearly tracks the H profile because CH$_3$ is a direct consequence of the oxidation of methane in R60}
\end{figure}

\begin{figure}
  \centering
    \includegraphics[width=0.5\textwidth]{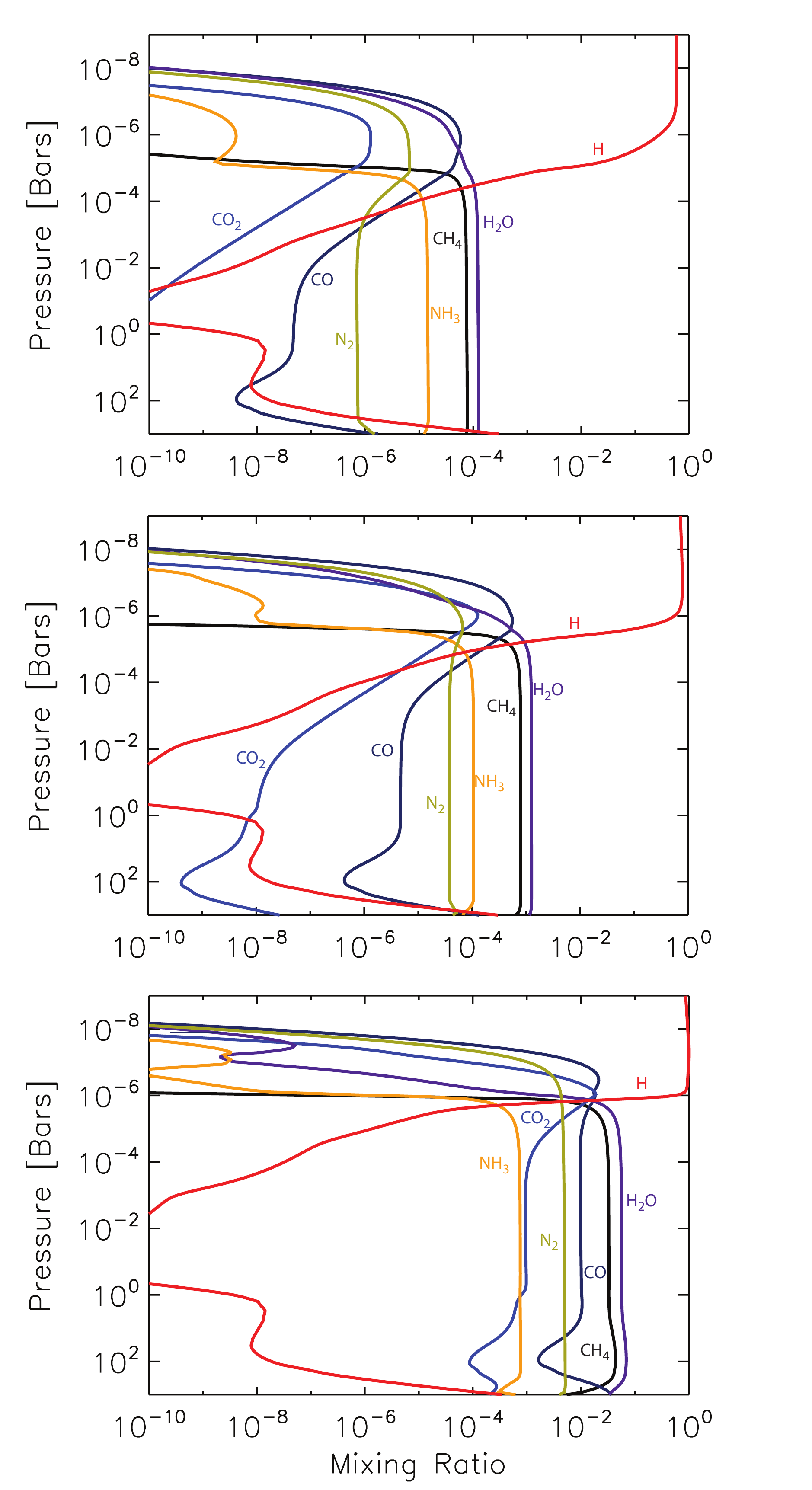}
     \caption{Photochemically derived mixing ratios for the abundant species (H$_2$O, CH$_4$, NH$_3$, N$_2$, CO, CO$_2$, and H) for $\zeta=0.1$ (top), $\zeta=1$ (middle), and $\zeta=50$ (bottom).}
\end{figure}

\begin{figure}
  \centering
    \includegraphics[width=0.5\textwidth]{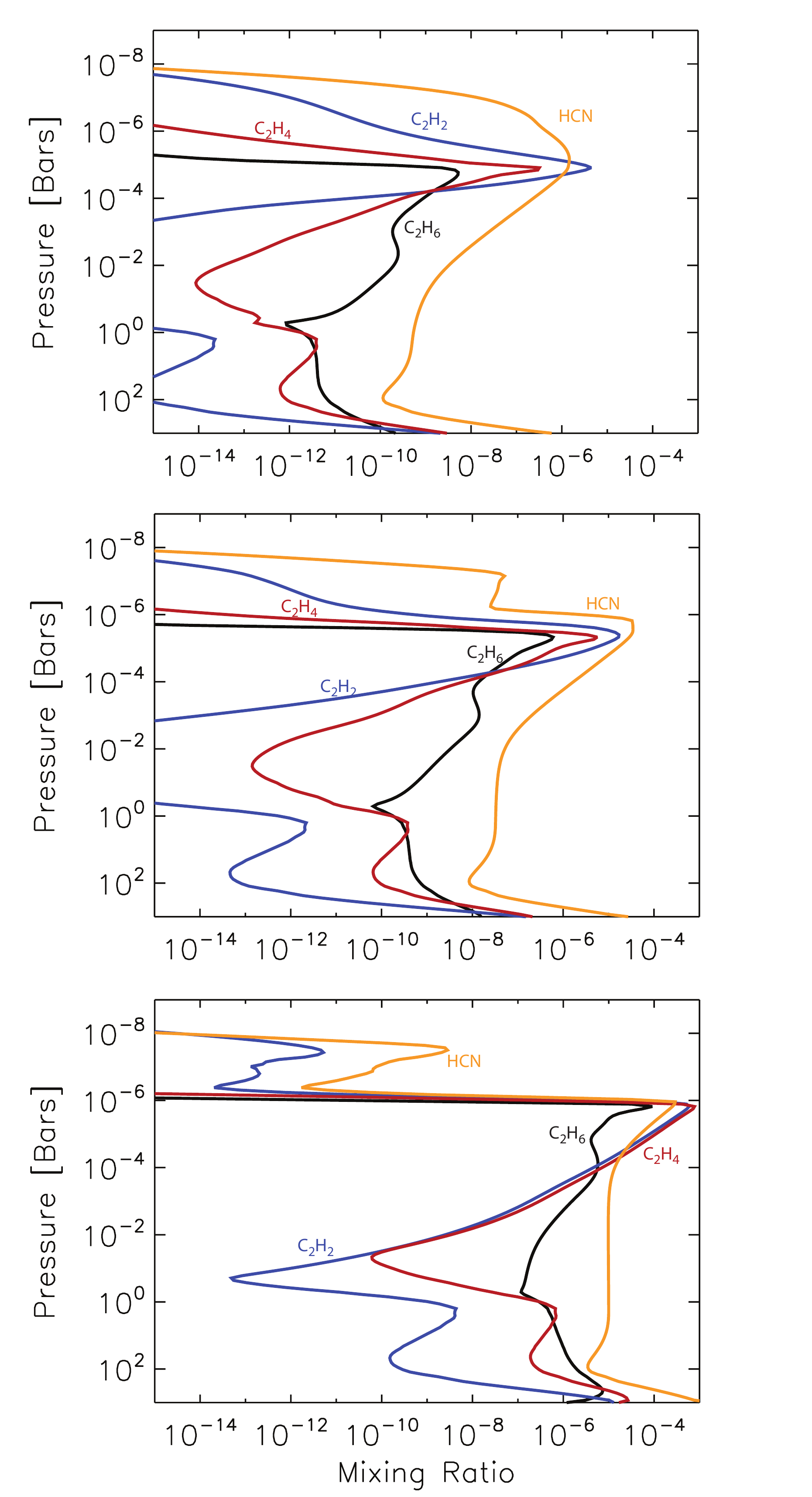}
     \caption{Photochemically derived mixing ratios for the disequilibrium species, the hydrocarbons and hydrogen cyanide, for $\zeta=0.1$ (top), $\zeta=1$ (middle), and $\zeta=50$ (bottom) }
\end{figure}

\subsubsection{Atomic H \& H$_{2}$O}
Arguably, the most important radical in these atmospheres is atomic
hydrogen.  Its relatively large abundance ($\sim$75\% above 1
$\mu$bar, Figure 6) drives the bulk of disequilibrium chemistry in the
upper atmosphere.  As is seen in Figures 5-7, when the atomic H
abundance increases with altitude, the concentration of disequilibrium species
increases with it.  Hydrogen attacks the
large stable reservoirs, NH$_3$ and CH$_4$, to build these
desequilibrium species.  In the cold Solar System giants, atomic
hydrogen is primarily produced by the photosensitized dissociation of
H$_{2}$ via heavier hydrocarbons, and the photodissociation of CH$_{4}$
and ethylene C$_2$H$_4$.  In hotter giant planets, as in GJ~436b, the atomic hydrogen is made
primarily by the photodissociation of water (Liang et
al., 2003, Line et al. 2010, Moses et al., 2010). This is because, unlike in the Solar
System's giants, water is not sequestered in clouds and is readily available for
photolysis.  Its large UV cross section combined with a large
thermochemical abundance, makes water the most important source of atomic
hydrogen in GJ436b.  The detailed mechanism for producing H
is the photosensitization of H$_{2}$ using water via,
\begin{center}
$$\rm{H_2O + h\nu \rightarrow OH + H } \eqno{R25}$$
$$\rm {H_2 + OH   \rightarrow H_2O + H } \eqno{R169}$$
\rule{2.5in}{1pt}
$$\rm {Net: H_2  \rightarrow 2H } \eqno{III}.$$
\end{center}
This photosensitization is efficient because
H$_{2}$O dissociates out to $\sim$2000 \AA, whereas H$_{2}$
dissociates only out to $\sim$800 \AA.  H$_{2}$O acts as a photon sink, with
factor $\sim 10^{4}$ more photons available for its photolysis,  than for direct H$_{2}$
photolysis. Because of these factors the net photosensitized destruction of H$_{2}$
by H$_2$O proceeds 5 orders-of-magnitude faster than the direct
photolysis of H$_{2}$, and 3 orders-of-magnitude faster than the
photosensitized destruction of H$_{2}$ via the hydrocarbons.  The
mixing ratio of water itself is largely unaltered below 1$\mu$bar levels.  

\subsubsection{CH$_{4}$ \& Hydrocarbons}

Thermochemically, methane is the most abundant
hydrocarbon. Overall it is the fourth most abundant species after H$_2$O,
H$_2$ and He, and it is the
parent molecule for the synthesis of all other hydrocarbons. Methane mixing ratios are
$\ge 10^{-4}$ at altitudes below the 0.1 mbar level, even
for the lowest metallicities. The models generally have methane mixing ratios at least 
3 orders-of-magnitude higher than concentrations retrieved from
the observations (Madhusudhan \& Seager 2011). 
Although photolysis seems not to
significantly modify methane abundances, it does produce
large concentrations of the methyl radical, CH$_3$; this radical is important in the synthesis of
heavier hydrocarbons.  CH$_3$ is formed by photosensitized
dissociation of methane.  The free atomic hydrogen from scheme $III$
readily attacks methane to produce H$_2$ and CH$_3$. The trigger and
pathway for this is:
\begin{center}
$$\rm{H_2O + h\nu \rightarrow OH + H}  \eqno{R25}$$
$$\rm{H_2 + OH  \rightarrow H_2O + H}    \eqno{R169}$$
$$\rm{H+ CH_4 \rightarrow  CH_3 + H_2} \eqno{R60}$$
\rule{2.5in}{1pt}
$$\rm{Net: CH_4  \rightarrow CH_3 + H}. \eqno{IV}$$
\end{center}
The methyl radical's mixing ratios can be as high as $\sim10^{-4}$, as in the $\zeta = 1$
case (Figure 5).  Due to the warmer upper atmosphere,
relative to that in the solar system giants, the oxidation of methane (via R60)
is more two orders-of-magnitude more efficient than direct photolysis.  
Because the forward reaction (R60) proceeds more sharply with rising temperature than
the reverse (R61), hotter upper atmospheres (as in
HD~189733b and HD~209448b) will have a tendency to destroy methane more readily, especially
when there are large quantities of photochemically produced atomic
hydrogen present.  This photosensitized destruction of methane causes it to decline
sharply above $\sim$10 $ \mu$bars; this is well below the planetary homopause, but well 
above the infrared photosphere (Figure 8).  It also drives the production of heavier
hydrocarbons.  Little to no heavier hydrocarbon (C$_n$H$_m$, where
$n,m \ge 2$) is expected via vertical mixing alone, with mixing ratios remaining below $\sim10^{-10}$ at altitudes above 1 bar.
Methane photosensitization (scheme IV) converts the carbon into ethylene
(C$_2$H$_4$), acetylene (C$_2$H$_2$), and ethane (C$_2$H$_6$) via
\begin{center}
$$\rm{H_2O + h\nu \rightarrow OH + H} \eqno{R25}$$
$$\rm{H_2 + OH \rightarrow H_2O + H } \eqno{R169}$$
$$\rm{2(H + CH_4\rightarrow CH_3 + H_2)} \eqno{R60}$$
$$\rm{CH_3 + CH_3 + M \rightarrow C_2H_6 + M} \eqno{R613}$$
$$\rm{H + C_2H_6 \rightarrow C_2H_5 + H_2} \eqno{R70}$$
$$\rm{H + C_2H_5 \rightarrow C_2H_4 + H_2} \eqno{R68}$$
$$\rm{H + C_2H_4 \rightarrow C_2H_3 + H_2} \eqno{R85}$$
$$\rm{H + C_2H_3 \rightarrow C_2H_2 + H_2} \eqno{R64}$$
\rule{2.5in}{1pt}
$$\rm{ Net: 2CH_4 + 4H \rightarrow C_2H_2 +5 H_2} \eqno{V}$$
\end{center}
The net reaction ultimately produces C$_2$H$_2$, making it the
most abundant heavy hydrocarbon.  This scheme is different than the solar system gas giants where the most dominant pathway for producing acetylene involves the binary collision between two $^3$CH$_{2}$ radicals.  This difference can again, be owed to the overwhelming abundance of atomic H from water photolysis which can readily reduce the ethane produced R613 to acetylene.   Over the range of metallicities considered ($\zeta = 0.1$
to 50), the peak values of C$_2$ hydrocarbons occur 
between 10 and 1 $\mu$bars. These mixing ratios of C$_2$H$_4$, C$_2$H$_2$, C$_2$H$_6$ lie
between  3$\times$10$^{-7}$-6$\times$10$^{-6}$,
5$\times$10$^{-6}$-4$\times$10$^{-4}$, and
5$\times$10$^{-9}$-6$\times$10$^{-5}$ (Figure 7; for integrated columns see
Table 1).  For comparison, the
peak values in Jupiter are, respectively, $\sim 2\times 10^{-6}$,
$5 \times 10^{-6}$, and $20 \times 10^{-6}$ (Moses et al. 2005).  In the Solar
System's giant planets, ethylene, acetylene, and ethane have strong
mid-infrared stratospheric emission features at 10.5, 13.7 and 12.1
$\mu$m respectively.   These
C$_2$ species can lead to further synthesis of higher order hydrocarbons
that can form hydrocarbon aerosols (Zahnle et al. 2009).  However,
the vapor pressures for these species are high (many bars) at these
temperatures, so it may be difficult to form such aerosols.
Additionally, Moses et al. 1992 showed that supersaturation ratios of
10 to 1000s may be required in order to trigger condensation due to
the lack of nucleation particulates in Jovian type atmospheres.
\begin{figure*}
  \centering
    \includegraphics[width=0.9\textwidth]{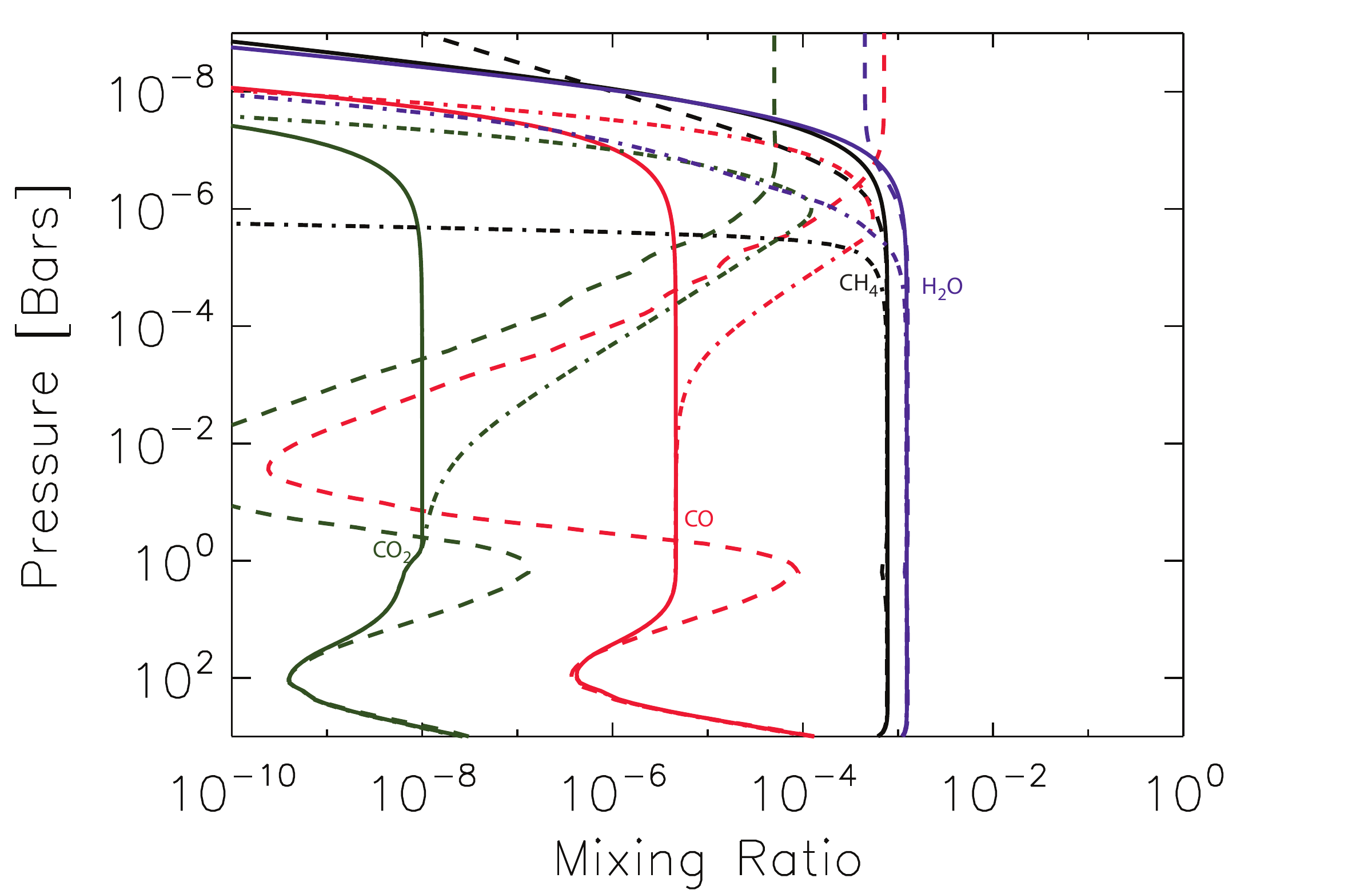}
     \caption{The effects of vertical mixing and photochemistry compared with the thermochemcial equilibrium profiles for methane, water, carbon monoxide, and carbon dioxide under solar abundances ($\zeta$=1).   The dashed curves are the thermochemcially derived mixing ratios (Figure 2 middle panel).  The solid curves are the mixing ratios with eddy mixing (as in Figure 3 middle panel).  The dot-dashed curves are the photochemical mixing ratios (Figure 5 middle panel). Note that methane becomes photochemically depleted near 1 $\mu$bar when compared to just vertical mixing (solid black curve).   CO and CO$_2$ are photochemically enhanced above the 10 mbar level. }
\end{figure*}
detectability

\subsubsection{CO \& CO$_{2}$}
As described in \S3.2, the CO abundance above 10 bars is
determined by the reaction rate of scheme I, and the strength of
vertical mixing.  In the absence of incident stellar UV, a
profile with a constant vertical mixing ratio up to the homopause is obtained.  With
incident UV radiation, there is a photochemical enhancement of CO
near the 1 $\mu$bar level, of up to a factor of $10^2$ for the $\zeta = 1$ case
(Figure 6, 8).  This high altitude enhancement is a property of
the cooler atmosphere of GJ~436b; in hot Jupiter atmospheres, 
as in HD~189733b and HD~209458b, such enhancements
or deficits will tend to be driven back towards equilibrium 
values.  The carbon in this extra CO is ultimately derived from the
CH$_4$ reservoir, via the following reaction scheme:
\begin{center}
$$\rm{H_2O + h\nu \rightarrow OH + H} \eqno{R25}$$
$$\rm{H_2 + OH   \rightarrow  H_2O + H} \eqno{R169}$$
$$\rm{H + CH_4  \rightarrow CH_3 + H_2} \eqno{R60}$$
$$\rm{H_2O + h\nu \rightarrow   O + 2H} \eqno{R26}$$
$$\rm{O + CH_3 \rightarrow  H_2CO + H_2} \eqno{R98}$$
$$\rm{H + H_2CO  \rightarrow HCO + H_2} \eqno{R233}$$
$$\rm{H + HCO \rightarrow  CO + H_2}  \eqno{R213}$$
\rule{2.5in}{1pt}
$$\rm{Net: H_2O + CH_4 \rightarrow  CO +2H_2 + 2H}\eqno{VI}$$
\end{center}
Scheme VI is driven by the water photolysis driven dissociation of 
CH$_4$ to CH$_3$ 
via scheme IV. Atomic O is produced by
photolytic fragmentation of water (R26); the net absorption cross
section for this branch is $\simeq 0.1$ that of the main
branch in R25.  The two radicals, O and CH$_3$, form formaldehyde in 
R98, and followed thereafter by a two-step conversion to CO (R233 and R213).  
An enhancement of CO$_2$ largely traces the enhancement of CO via:
\begin{center}
$$\rm{H_2O + h\nu \rightarrow OH + H }\eqno{R25 }$$
$$\rm{OH + CO  \rightarrow  CO_2 + H }\eqno{R187}$$
\rule{2.5in}{1pt}
$$\rm{Net: H_2O + CO \rightarrow CO_2 + 2H}\eqno{VII}$$
\end{center}
Photochemically enhanced CO$_2$ mixing ratios reach $\sim 10^{-4}$  at
1 $\mu$bar for $\zeta = 1$.  Column averaged mixing ratios are
$5\times10^{-6}$ and $6\times10^{-9}$ (see Table 1).  This is low
compared to the observed mixing ratios of $\sim$1$\times$10$^{-4}$ and
$\sim$1$\times$10$^{-7}$, respectively.  Increasing the metallicity to
$\zeta = 50$, increases the mixing ratios to $\sim$1$\times$10$^{-2}$ and
$\sim$5$\times$10$^{-4}$, suggesting that the observed CO
and CO$_2$ columns are consistent with a metallicity enhanced to levels  
observed in Solar System's ice giant planets (Table 1). 

\subsubsection{Nitrogen \& HCN}

Ammonia and molecular nitrogen, N$_2$, are thermochemically the two
most stable species in a reducing atmosphere and their relative abundance
within the $1 - 0.001$ bar pressure levels is dictated by quench chemistry.
Because it is relatively abundant, the addition of hot (quenched or otherwise) NH$_3$ 
(Tennyson et al. 2010) to the list of
absorbers used for model fitting and retrieval may well be quite important.
Other important N species are mainly photochemical
byproducts, with HCN being the most abundant photochemically produced
molecule between 1 and 0.1 mbar levels, having mixing ratios of typically $10^{-6}$ ($\zeta = 1$)
to $10^{-5}$ ($\zeta = 50$) at 0.1 mbar. Peak HCN occurs well above the photospheric levels, 
approaching $10^{-4}$ at 1 $\mu$bar.
The synthesis of HCN is initiated via water and ammonia photolysis, and completed by
subsequent reactions between the ammonia and methane derived radicals: 
\begin{center}
$$\rm{H_2O + h\nu  \rightarrow OH + H}\eqno{R25} $$
$$\rm{ H_2 + OH\rightarrow H_2O + H} \eqno{R169} $$
$$\rm{H + CH_4 \rightarrow CH_3 + H_2}\eqno{R60} $$
$$ \rm{ NH_3 + h\nu \rightarrow NH_2 + H}\eqno{R43} $$
$$\rm{ H + NH_2 \rightarrow NH + H_2} \eqno{R455} $$
$$\rm{NH + CH_3 \rightarrow CH_2NH + H} \eqno{R685} $$
$$\rm{H + CH_2NH  \rightarrow H_2CN + H_2} \eqno{R655} $$
$$\rm{H + H_2CN \rightarrow HCN + H_2} \eqno{R663} $$
\rule{2.5in}{1pt}
$$\rm{Net: CH_4+NH_3\rightarrow HCN + 3H_2}\eqno{VIII} $$
\end{center}
We note that R43, the photolysis of ammonia to amino radical, is the
most important pathway for  NH$_2$ formation at pressures greater than
10 $\mu$bar. At lower pressures this reaction is driven by ammonia
photosensitzation,
$$ \rm{ NH_3 + H \rightarrow NH_2 + H_2},  \eqno{R454} $$
where the is H derived from H$_2$O photolysis. 
In conclusion when water, ammonia and methane are
present, disequilibrium HCN is relatively abundant.
The best chance for the detection of HCN is via the transmission spectroscopy of
its vibrational fundamental
bands at 3 and 14 $\mu$m (Shabram et al. 2011).

\subsubsection{Sulfur}
\begin{figure}
  \centering
    \includegraphics[width=0.5\textwidth]{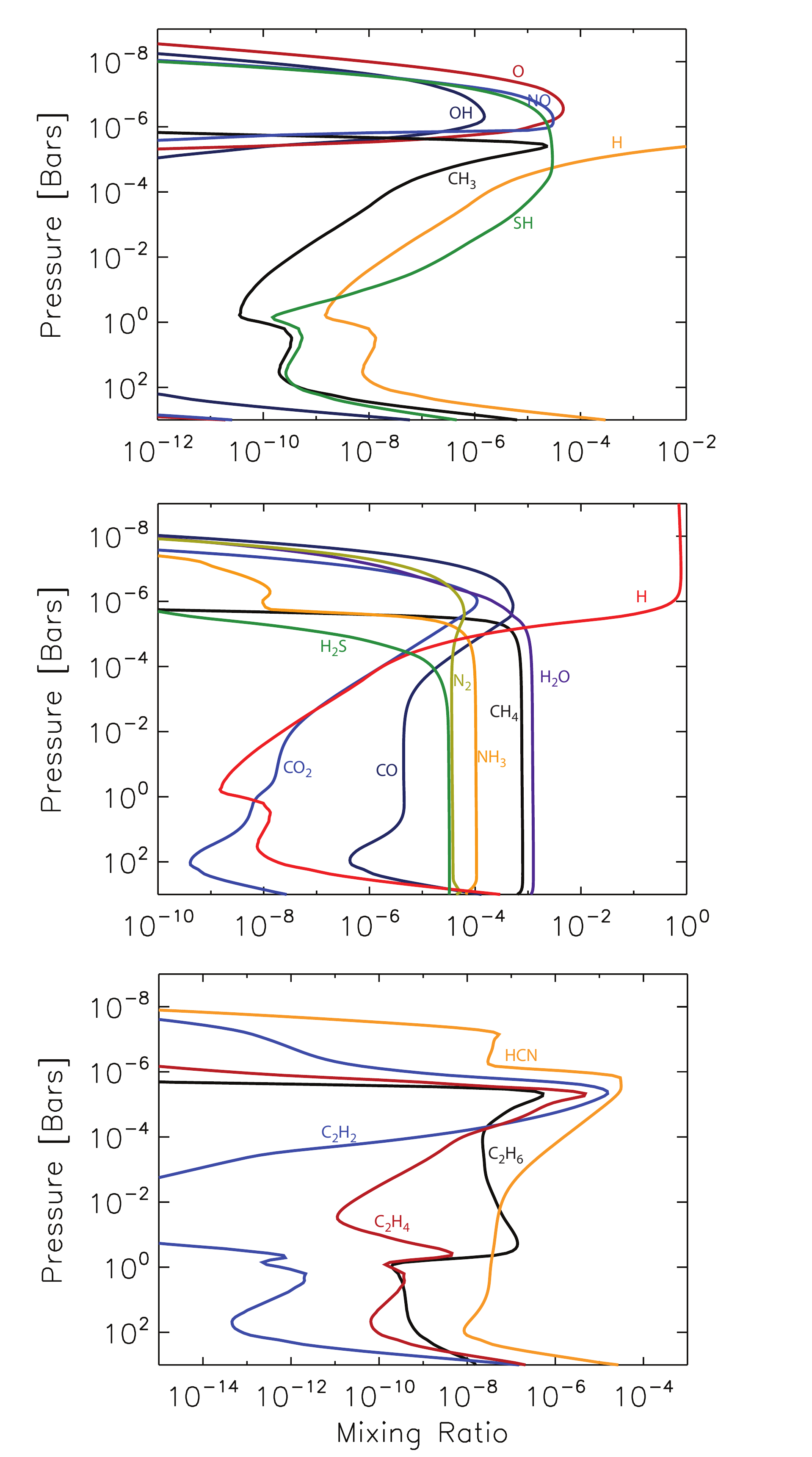}
     \caption{Photochemically derived mixing ratios for $\zeta=1$ in the presence of sulfur species (H$_2$S and HS) for the radicals (top), abundant species (middle), and disequilibrium species (bottom).   Compare this Figure to the $\zeta=1$  cases in Figures 5, 6, and 7 to see the effects of H$_2$S on the mixing ratios.  Note that the abundance of atomic H is enhanced by orders of magnitude between 1 and 10$^{-4}$ bars as a result of scheme IX.  This H increase enhances the hydrocarbon abundances significantly in this portion of the atmosphere.}
\end{figure}

Because atomic H attacks both CH$_4$ and NH$_3$, we examine the role
of H$_2$S as a source of free H (Zahnle et al. 2009); 
S is isoelectronic with and similar in chemical properties to O, but has a considerably reduced primordial abundance, with S/O $\simeq 0.02$. 
In a subset of models, we introduce the following (very restricted) set of sulfur reactions
with accurate laboratory determined reaction rates:
\begin{center}
$$\rm{H_2S + h\nu  \rightarrow SH + H}\eqno{R705 ~~~~~~~~} $$
$$\rm{H_2S + H \rightleftharpoons SH + H_2} \eqno{R701,~R702} $$
$$\rm{H_2S + OH \rightleftharpoons SH + H_2O} \eqno{R703,~R704} $$
\end{center}
H$_2$S is an attractive source of free hydrogen due to
its ability to photodissociate out to relative long wavelengths, $\sim$2600~\AA. It has a photolysis rate constant comparable 
to that of H$_2$O, and we 
find a $10^2$ enhancement in H between the pressure levels of
1 bar and 0.1 mbar upon including these two sulfur species (Figure 9); the relevant
reactions are:
\begin{center}
$$\rm{H_2S + h\nu  \rightarrow SH + H}\eqno{R715} $$
$$\rm{SH + H_2  \rightarrow H_2S+H}\eqno{R712} $$
\rule{2.5in}{1pt}
$$\rm{Net: H_2\rightarrow 2H}\eqno{IX} $$
\end{center}
This enhanced H abundance is catalyzed by the photolysis of
H$_2$S (traced by the SH radical in figure 9, top panel). The atomic H reacts efficiently with CH$_4$ in R60,
producing an increased concentration of the radical CH$_3$, which in turn 
drives hydrocarbon production (scheme V) near the 0.1 bar
level.  

However, the free H in the middle atmosphere, does little to affect the CH$_4$ mixing ratios; this is because 
the S/C abundance ratio is low.
Sulfur would need to be enriched by a substantial factor of $\sim$20, over the solar S/C value, in order for 
H$_2$S to have an appreciable impact on atmospheric CH$_4$.  
Although the few considered sulfur species (H$_2$S, SH) do not much impact the overall chemistry,
it is possible that another sulfur compound, such as SO, may act as a
catalyst assisting in the conversion of reduced carbon into oxidized
carbon.  Previously, Moses (1996) has modeled the SL9 Jupiter impact and
shown the importance of S in many reaction schemes involving both
C and N species, and so the role of S chemistry in the hot
extrasolar giants should continue to be investigated in the future (see Zahnle et al. 2009).

\begin{center}
\begin{deluxetable*}{ccccccccc}
\tabletypesize{\footnotesize}
\tablecolumns{6} 
\tablewidth{0pt} 
\tablecaption{Photochemical model abundances compared with the observations of Stevenson et al. 2010 (S10), Madhusudhan \& Seager (2011) (MS10) and Beaulieu et al. (2010)  (B10).  The model abundances are the integrated column mixing ratios between 7 bars and 0.1 bars, the pressure levels sampled by the observations, for 0.1$\times$, 1$\times$, and 50$\times$ solar elemental abundances. 
      }
\tablehead{\colhead{Molecule }			&
		\colhead {0.1$\times$ }                                    &
                 \colhead{1$\times$ }                             &
                \colhead{50$\times$ }                   		&
                 \colhead{MS10 }                             &
                 \colhead{S10 }                             &
                  \colhead{ B10}                             &         
                }
\startdata
CH$_4$&7.66 $\times$10$^{-05}$ &7.90$\times$10$^{-04}$ &2.96 $\times$10$^{-02}$&(3-6)$\times$10$^{-06}$ &1$\times$10$^{-07}$ &5$\times$10$^{-04}$\\
CO & 4.22$\times$10$^{-08}$&4.29$\times$10$^{-06}$ &8.56$\times$10$^{-03}$&(3-100)$\times$10$^{-05}$ & (1-7)$\times$10$^{-04}$&--\\
CO$_2$&7.74$\times$10$^{-12}$&6.09$\times$10$^{-09}$& 5.44$\times$10$^{-04}$&(1-10)$\times$10$^{-07}$ &(1-10)$\times$10$^{-07}$&--\\
H$_2$O&1.25$\times$10$^{-04}$&1.26$\times$10$^{-03}$& 5.09$\times$10$^{-02}$& $\le$1$\times$10$^{-03}$ &(3-100)$\times$10$^{-06}$&--\\
HCN&4.84$\times$10$^{-10}$&3.09$\times$10$^{-08}$ &8.41 $\times$10$^{-06}$&-- & --&--\\
C$_2$H$_2$&1.21$\times$10$^{-14}$&1.18$\times$10$^{-12}$ &2.10 $\times$10$^{-09}$&--& --&--\\
NH$_3$&1.45$\times$10$^{-05}$&1.06$\times$10$^{-04}$ & 6.54$\times$10$^{-04}$&-- & --&--\\
H$_2$S&--& $3.22\times$10$^{-05}$&--&-- & --&--\\
		
\enddata
\end{deluxetable*}
\end{center}

\section{Discussion \& Conclusions}
We have developed a 1D ``thermochemical and photochemical kinetics with
transport'' model following Visscher et al. (2010) and recently, Moses et al. (2011)
for extrasolar planet atmospheres. We use a
compilation of bidirectional reactions of the five most abundant
elements to model both the equilibrium and disequilibrium portions
of the atmosphere.
Using detailed balance with both forward and reverse reactions,
allows our model to reach thermochemical equilibrium
kinetically, thereby obviating the need to choose {\it ad hoc} lower  
boundaries for
multiple quenched species, and allowing a seamless transition between the transport
dominated and the chemical equilibrium zones. A limitation is that  
we adopt a static temperature structure;
a future improvement would allow the iterative adjustment and co-evolution of the temperature
structure with the chemistry. Also, the eddy diffusivity
profile $K_{zz}(z)$ is poorly constrained, and is
essentially a free parameter in any of these models.

We have applied our models to study the atmosphere of the
transiting Neptune-like planet GJ~436b. The elemental abundance of
atmosphere, a key input parameter, is relatively
uncertain, but mass-radius constraints suggest that GJ~436b
must be enriched to at least $10\times$ solar levels. We model a range
of atmospheric enrichment to cover this
instrinsic uncertainty; we observe the trends when varying
$\zeta$, and rule out the possibility that intermediate values of
$\zeta$ would spring any surprises. The UV fluxes of stars
other than the Sun are often difficult to obtain. M-dwarf hosts
can be chromospherically hyperactive, and because UV photolysis may
drive the depletion of weakly bonded molecules
such as CH$_4$, NH$_3$ and H$_2$S, it is
important to have an accurate UV estimate for GJ~436. We use a
combination of GALEX and HST UV fluxes along with Rosat and XMM-Newton soft X-ray fluxes
to bound the UV continuum and line emission of GJ~436.

The GJ~436b model atmospheres show that a combination of
photochemistry, chemical kinetics and transport-induced quenching drives
the composition well out of equilibrium. While equilibrium conditions
are maintained in the deep, hot, troposphere (below a 10s of bars for
$\rm CO\rightleftharpoons CH_4$, and 100s of bars for
$\rm N_2\rightleftharpoons NH_3$), the composition of the middle atmosphere is altered
by the dredging up of quenched gases such as CO and NH$_3$. The effects of
transport disequilibrium are prominent in cooler planets such as GJ~436b
because the quench points for major species depend on the temperature.  
As it gets colder, the pressure points for quenching are pushed deeper into
the atmosphere due to the longer interconversion timescales from one species reservoir to another.  In contrast to the quenched species  
(CO, CO$_2$, NH$_3$),
the effect of vertical mixing on the reservoir gases such as
CH$_4$ and H$_2$O is relatively feeble.

The reservoir gases H$_2$O and CH$_4$, and NH$_3$ are largely unaffected by 
photochemistry because of their (a) large abundances, and (b) rapid recycling. 
Nevertheless, it is their photolysis that drives the bulk of the disequilibrium chemistry in the 
upper atmosphere producing CH$_4$ and NH$_3$ sinks such as 
heavier hydrocarbons (such as C$_2$H$_2$, etc.) and simple nitriles (such as HCN). 
Much as in the hot Jupiters (Liang et al. 2003), H is the most  
important and active atom
in the bulk of the atmosphere;  it is created by the photosensitized
destruction of H$_2$, catalyzed by the presence of H$_2$O and H$_2$S. The  
latter gas, though less
abundant than water, is important because of its ability to capture  
incident starlight photons
out wavelengths as long as  $2600$ \AA. In most models, H replaces H$_2$ as the most abundant
species in the atmosphere above the planetary homopause at $\simeq 1$ $ 
\mu$bar. Because CH$_4$ is the largest C carrier in the planet's UV photosphere, we  
create abundant
C$_2$ compounds (Figure 7) despite the relatively efficient hydrogenation back to CH$_4$.
Species such as acetylene, C$_2$H$_2$, formed in abundance in our enriched models, are precursors for potential 
hydrocarbon soot
formation in the upper atmosphere (as opposed to the hotter Jupiters such as HD~209458b and  
HD~189733b, wherein CO carries the bulk of carbon in the stratosphere). 
Our reaction lists for hyrdocarbon chemistry are truncated at C$_2$, and so we do not synthesize C$_3$ and 
heavier hydrocarbons and nitriles explicitly.

Within the range physical and chemical processes captured in our models, and 
the considered reaction sets and their kinetics, we find it difficult explain the observations 
suggesting a methane-poor
GJ~436b. Except above 1 $\mu$bar pressure levels where CH$_4$
is photochemically converted to CO, HCN and C$_2$ hydrocarbons,  it remains the predominant C 
reservoir in the lower atmosphere and in the region of the IR photosphere. The
observed abundances of quenched CO and CO$_2$ are in agreement with an
atmosphere enriched to levels intermediate between 1 to 50 times solar (as in Madhusudhan
\& Seager 2011). The depleted water may either contrarily suggest a sub solar metallicity (Table 1), or skewed heavy metals
ratios; the latter is a possibility which we have not considered herein as there are far too many combinations to
explore. In the 1$\times$ solar models, the
methane abundance is consistent with the values retrieved by 
Beaulieu et al. (2010) (Table 1) using transit observations. We suppose
it is possible that a more complete inclusion of other relatively abundant elements such as
S and P, or distorted elemental ratios (C/O or O/S etc.), or ill-understood chemistry and
exotic processes (not considered herein, such as the 3 dimensionality of the problem) could do more
to explain the chemistry of this enigmatic atmosphere.

We agree with Moses et al. (2011) that quench level arguments can be used to predict abundances, so long as this is done with the 
appropriate level of caution.  By this, we mean that the relevant rate-limiting reaction must necessarily be identified in order to 
properly calculate the timescale for chemical loss. 
Also, quenched gases do not share a common quench level and assuming so can result in gross under- or overestimation of their 
abundances.   For example, as shown herein, N$_2$ and CO have vastly different quench levels.   
For the moderate to high levels of incident UV flux, photolysis 
generates high concentrations of secondary byproducts, but does not significantly alter the abundances 
of the reservoir gases; in our estimation photochemistry cannot alter the dayside methane budget. Hotter atmospheres with
sluggish vertical mixing and hot stratospheres are required for severe methane depletion. For example, in figure 10, we approximate
such as atmosphere as  isothermal with $T = 1200$ K,  $\zeta = 5$, and $K_{zz}$=$1\times10^{6}$ cm$^2$s$^{-1}$, and with zero
UV irradiation (similar to models by Zahnle et al. 2009).  In this hypothetical atmosphere there is relatively little quenched methane.  At $T = 1200$ K and low pressures, the rate determining step for CH$_4$ $\rightarrow$ CO (reverse of R351) is faster than the vertical transport time 
throughout the atmosphere, allowing the CH$_4$ to be in thermochemical equilibrium with CO everywhere (Figure 10). 
Since equilibrium conditions apply, the $P^2$ term in equation 2 results in the rapid vertical fall-off of CH$_4$.

  \begin{figure}
  \centering
    \includegraphics[width=0.5\textwidth]{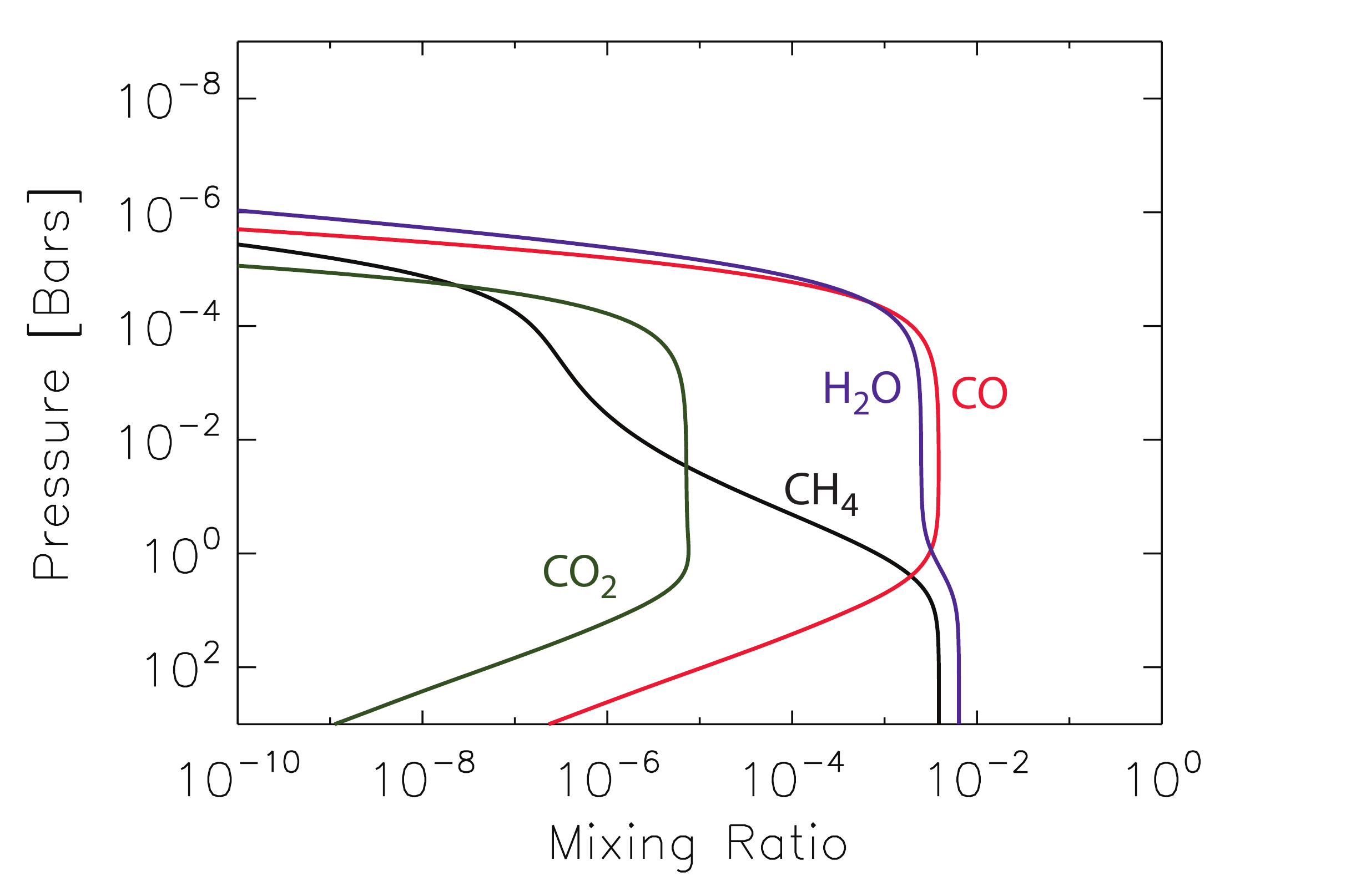}
     \caption{Mixing ratios for, CO, CH$_4$, CO$_2$, and H$_2$O  assuming a T=1200 K isothermal atmosphere, $\zeta=5 $, and $K_{zz}$=$1\times10^{6}$ cm$^2$s$^{-1}$, with no photochemistry.  The observed methane fall-off is due to high temperatures alone; high temperatures imply a short chemical loss time for of CH$_4$.  Because of the large transport time, CH$_4$ and CO are nearly in equilibrium at all altitudes.}
\end{figure}

The models presented herein are by no means restricted in applicability to GJ~436b like Neptunes, and much of the
modeled chemical state may be generalized to
H/He dominated planets in the 500-1000 K temperature
range.  In this regime CH$_4$ is the
primary carbon carrier and CO is quenched.  The reverse is true in hotter atmospheres, $T > 10^3$ K,
where CO is the primary carbon carrier and CH$_4$ is quenched.  
NH$_3$ is quenched deep in the atmosphere and can be quite abundant in the photosphere.  
Higher hydrocarbons and HCN are produced photochemically in relatively high abundances 
at mbar to $\mu$bar pressures. Similarly, an enhancement of   
CO and CO$_2$ over the quench concentrations, driven by the photolysis of H$_2$O, is observed in the high atmosphere.  
Water is in gaseous phase and abundant, and not condensed out as it would be in cooler atmospheres.  
GJ~1214b, a $T \simeq 500$ K low super Earth or mini Neptune, also orbiting an M dwarf primary
(Charbonneau et al. 2009; Sada et al. 2010; Bean et al. 2010; Carter et al. 2011; D{\'e}sert et al. 2011), falls in this regime
of warm atmospheres.
If GJ~1214b is in possession of a reducing H-He atmosphere
(Croll et al. 2011; Crossfield et al. 2011), much of the atmospheric chemistry would be analogous to that in GJ~426b; 
this, however, is speculative as there is much current debate over the bulk composition of GJ~1214b.

\section{Acknowledgements}		
We would like to thank Julie Moses, Channon Visscher, Karen Willacy, and M.C. Liang for useful chemistry discussions and tips.  We would also like to thank Xi Zhang, Heather Knutson,  Mimi Gerstell, Mark Allen, the  Yuk Yung Group, and the anonymous referee for reading the manuscript and providing valuable feedback.  M. Line is supported by the JPL Graduate Fellowship funded by the JPL Research and Technology Development Program.  P. Chen \& G. Vasisht are supported by the JPL Research \& Technology Development Program, and contributions herein were supported by the Jet Propulsion Laboratory, California Institute of Technology, under a contract with the National Aeronautics and Space Administration.


\begin{thebibliography}

\bibitem[Allen et al.(1981)]{1981JGR....86.3617A} Allen, M., Yung, Y.~L., \& Waters, J.~W.\ 1981, \jgr, 86, 3617 
\bibitem[Anders et al. (1982)]{1982} Anders, E. \& Ebihara, M., 1981, Geochem. Cosmo. Chem. Acta., 46, p. 2363
\bibitem[Asplund et al.(2009)]{2009ARA&A..47..481A} Asplund, M., Grevesse, N., Sauval, A.~J., \& Scott, P.\ 2009, \araa, 47, 481
\bibitem[Batygin et al.(2009)]{2009ApJ...699...23B} Batygin, K., Laughlin, G., Meschiari, S., Rivera, E., Vogt, S., \& Butler, P.\ 2009, \apj, 699, 23
\bibitem[Bean et al.(2006)]{2006ApJ...653L..65B} Bean, J.~L., Benedict, G.~F., \& Endl, M.\ 2006, \apjl, 653, L65 
\bibitem[Bean et al.(2010)]{2010Natur.468..669B} Bean, J.~L., Kempton, E.~M.-R., \& Homeier, D.\ 2010, \nat, 468, 669 
\bibitem[Beaulieu et al.(2011)]{2011ApJ...731...16B} Beaulieu, J.-P., et al.\ 2011, \apj, 731, 16
\bibitem[Butler et al.(2004)]{2004ApJ...617..580B} Butler, R.~P., Vogt, S.~S., Marcy, G.~W., Fischer, D.~A., Wright, J.~T., Henry, G.~W., Laughlin, G., \& Lissauer, J.~J.\ 2004, \apj,617, 580 
\bibitem[Carter et al.(2011)]{2011ApJ...730...82C} Carter, J.~A., Winn, J.~N., Holman, M.~J., Fabrycky, D., Berta, Z.~K., Burke, C.~J., \& Nutzman, P.\ 2011, \apj, 730, 82 
\bibitem[Charbonneau et al.(2009)]{2009Natur.462..891C} Charbonneau, D., et al.\ 2009, \nat, 462, 891 
\bibitem[Chase et al.(1985)]{1985J. Phys. Chem. Ref. Data}Chase Jr, M.W. \& Davies, C.\ 1985, J. Phys. Chem. Ref. Data, 14, 1856
\bibitem[Croll et al.(2011)]{2011arXiv1104.0011C} Croll, B., Albert, L., Jayawardhana, R., Miller-Ricci Kempton, E., Fortney, J.~J., Murray, N., 
\& Neilson, H.\ 2011, arXiv:1104.0011 
\bibitem[Crossfield et al.(2011)]{2011arXiv1104.1173C} Crossfield, I.~J.~M., Hansen, B.~M.~S., \& Barman, T.\ 2011, arXiv:1104.1173
\bibitem[Cooper \& Showman(2006)]{2006ApJ...649.1048C} Cooper, C.~S., \& Showman, A.~P.\ 2006, \apj, 649, 1048 
\bibitem[Doyle et al.(1997)]{1997A&A...318..835D} Doyle, J.~G., Mathioudakis, M., Andretta, V., Short, C.~I., \& Jelinsky, P.\ 1997, \aap, 318, 835 
\bibitem[Deming et al.(2007)]{2007ApJ...667L.199D} Deming, D., Harrington, J., Laughlin, G., Seager, S., Navarro, S.~B., Bowman, W.~C., \& Horning, K.\ 2007, \apjl, 667, L199 
\bibitem[D{\'e}sert et al.(2011)]{2011ApJ...731L..40D} D{\'e}sert, J.-M., 
et al.\ 2011, \apjl, 731, L40 
\bibitem[Ehrenreich et al.(2011)]{2011A&A...529A..80E} Ehrenreich, D., Lecavelier Des Etangs, A., \& Delfosse, X.\ 2011, \aap, 529, A80 
\bibitem[Figueira et al.(2009)]{2009A&A...493..671F} Figueira, P., Pont, F., Mordasini, C., Alibert, Y., Georgy, C., \& Benz, W.\ 2009, \aap, 493, 671
\bibitem[Gillon et al. (2007)]{2007A&A...472L..13G} Gillon, M., Pont, F., Demory, B.~O., Mallmann, F., Mayor, M., Mazeh, T., Queloz, D., Shporer,
A., Udry, S., \& Vuissoz, C.\ 2007, 472, L13  
\bibitem[Gizis et al.(2002)]{2002AJ....123.3356G} Gizis, J.~E., Reid, I.~N., \& Hawley, S.~L.\ 2002, \aj, 123, 3356 
\bibitem[Gladstone et al.(1996)]{1996Icar..119....1G} Gladstone, G.~R., Allen, M., \& Yung, Y.~L.\ 1996, Icarus, 119, 1 
\bibitem[Gordon \& McBride.(1996)]{1996 CEA}Gordon, S., McBride, B.J., \& NASA Tech. Info. Program, \ 1996, Computer program for calculation of complex chemical equilibrium compositions and applications, National Aeronautics and Space Administration, Office of Management, Scientific and Technical Information Program
\bibitem[Griffith \& Yelle(1999)]{1999ApJ...519L..85G} Griffith, C.~A., \& Yelle, R.~V.\ 1999, \apjl, 519, L85 
\bibitem[Gurvich et al. (1989)]{1989 Thermo}Gurvich, L.V. and Veyts, IV and Alcock, CB, institut prikladno khimii (Soviet Union) \ 1989, Thermodynamic properties of individual substances, Hemisphere Publisher
\bibitem[Hubeny 
\& Burrows(2007)]{2007ApJ...669.1248H} Hubeny, I., \& Burrows, A.\ 2007, \apj, 669, 1248 
\bibitem[H{\"u}nsch et al.(1999)]{1999A&AS..135..319H} H{\"u}nsch, M., Schmitt, J.~H.~M.~M., Sterzik, M.~F., \& Voges, W.\ 1999, \aaps, 135, 319 
\bibitem[Jodkowski et al. (1999)]{1999 J. Phys. Chem.A}Jodkowski, J.T. and Rayez, M.T. and Rayez, J.C. and B{\'e}rces, T. and D{\'o}b{\'e}, S. \ 1999, J. Phys. Chem. A., 103, 3750
\bibitem[Knutson et al.(2009)]{2009ApJ...690..822K} Knutson, H.~A., et al.\ 2009, \apj, 690, 822 
\bibitem[Kramm et al.(2011)]{2011A&A...528A..18K} Kramm, U., Nettelmann, N., Redmer, R., \& Stevenson, D.~J.\ 2011, \aap, 528, A18 
\bibitem[Landsman \& Simon(1993)]{1993ApJ...408..305L} Landsman, W., \& Simon, T.\ 1993, \apj, 408, 305 
\bibitem[Lewis et al.(2010)]{2010ApJ...720..344L} Lewis, N.~K., Showman, A.~P., Fortney, J.~J., Marley, M.~S., Freedman, R.~S., \& Lodders, K.\ 2010, \apj, 720, 344 
\bibitem[Liang et al.(2003)]{2003ApJ...596L.247L} Liang, M.-C., Parkinson, C.~D., Lee, A.~Y.-T., Yung, Y.~L., \& Seager, S.\ 2003, \apjl, 596, L247
\bibitem[Liang et al.(2004)]{2004ApJ...605L..61L} Liang, M.-C., Seager, S., Parkinson, C.~D., Lee, A.~Y.-T., \& Yung, Y.~L.\ 2004, \apjl, 605, L61 
\bibitem[Line et al.(2010)]{2010ApJ...717..496L} Line, M.~R., Liang, M.~C., \& Yung, Y.~L.\ 2010, \apj, 717, 496 
\bibitem[Lodders(2002)]{2002ApJ...577..974L} Lodders, K.\ 2002, \apj, 577, 974 
\bibitem[Mardling(2008)]{2008arXiv0805.1928M} Mardling, R.~A.\ 2008, arXiv:0805.1928 
\bibitem[Moses(1996)]{1996ccsl.proc..243M} Moses, J.~I.\ 1996, IAU Colloq.~156: The Collision of Comet Shoemaker-Levy 9 and Jupiter, 243 
\bibitem[Moses et al.(1992)]{1992Icar...99..318M} Moses, J.~I., Allen, M., \& Yung, Y.~L.\ 1992, Icarus, 99, 318 
\bibitem[Moses et al.(2005)]{2005JGRE..11008001M} Moses, J.~I., Fouchet, T., B{\'e}zard, B., Gladstone, G.~R., Lellouch, E., \& Feuchtgruber, H.\ 2005, Journal of Geophysical Research (Planets), 110, 8001 
\bibitem[Moses et al.(2011)]{2011arXiv1102.0063M} Moses, J.~I., et al.\ 2011, arXiv:1102.0063 
\bibitem[Nettelmann et al.(2010)]{2010A&A...523A..26N} Nettelmann, N., Kramm, U., Redmer, R., \& Neuh{\"a}user, R.\ 2010, \aap, 523, A26 
\bibitem[Prinn \& Barshay(1977)]{1977Sci...198.1031P} Prinn, R.~G., \& Barshay, S.~S.\ 1977, Science, 198, 1031 
\bibitem[Ribas et al.(2008)]{2008ApJ...677L..59R} Ribas, I., Font-Ribera, A., \& Beaulieu, J.-P.\ 2008, \apjl, 677, L59 
\bibitem[Rogers \& Seager(2010)]{2010ApJ...716.1208R} Rogers, L.~A., \& Seager, S.\ 2010, \apj, 716, 1208 
\bibitem[Sada et al.(2010)]{2010ApJ...720L.215S} Sada, P.~V., et al.\ 2010, \apjl, 720, L215 
\bibitem[Sanz-Forcada et al.(2010)]{2010ASPC..430..530S} Sanz-Forcada, J., Garc{\'{\i}}a-{\'A}lvarez, D., Velasco, A., Solano, E., Ribas, I., Micela, G., \& Pollock, A.\ 2010, Pathways Towards Habitable Planets, 430, 530 
\bibitem[Shabram et al.(2011)]{2011ApJ...727...65S} Shabram, M., Fortney, J.~J., Greene, T.~P., \& Freedman, R.~S.\ 2011, \apj, 727, 65 
\bibitem[Showman et al.(2009)]{2009ApJ...699..564S} Showman, A.~P., Fortney, J.~J., Lian, Y., Marley, M.~S., Freedman, R.~S., Knutson, H.~A., \& Charbonneau, D.\ 2009, \apj, 699, 564
\bibitem[Smith(1998)]{1998Icar..132..176S} Smith, M.~D.\ 1998, Icarus, 132, 176 
\bibitem[Stevenson et al.(2010)]{2010Natur.464.1161S} Stevenson, K.~B., et al.\ 2010, \nat, 464, 1161
\bibitem[Stothard et al.(1995)]{1995Chem}Stothard, N. and Humpfer, R. and Grotheer, H.H., \ 1995, Chem. Phys. Lett., 240, 474
\bibitem[Saumon et al.(2007)]{2007ApJ...656.1136S} Saumon, D., et al.\ 
2007, \apj, 656, 1136
\bibitem[Saumon et al.(2006)]{2006ApJ...647..552S} Saumon, D., Marley, 
M.~S., Cushing, M.~C., Leggett, S.~K., Roellig, T.~L., Lodders, K., 
\& Freedman, R.~S.\ 2006, \apj, 647, 552
\bibitem[Saumon et al.(2003)]{2003astro.ph.10805S} Saumon, D., Marley, 
M.~S., \& Lodders, K.\ 2003, arXiv:astro-ph/0310805
\bibitem[Swain et al.(2009)]{2009ApJ...690L.114S} Swain, M.~R., Vasisht, G., Tinetti, G., Bouwman, J., Chen, P., Yung, Y., Deming, D., \& Deroo, P.\ 2009, \apjl, 690, L114 
\bibitem[Tennyson(2010)]{2010epsc.conf..172T} Tennyson, J.\ 2010, European 
Planetary Science Congress 2010, held 20-24 September in Rome, 
Italy.~<A href=''http://meetings.copernicus.org/epsc2010''>http://meetings.copernicus.org/epsc2010</A>, p.172, 172 
\bibitem[Torres et al.(2008)]{2008ApJ...677.1324T} Torres, G., Winn, J.~N., \& Holman, M.~J.\ 2008, \apj, 677, 1324 
\bibitem[Visscher et al.(2010)]{2010Icar..209..602V} Visscher, C., Moses, J.~I., \& Saslow, S.~A.\ 2010, Icarus, 209, 602 
\bibitem[Walkowicz et al.(2008)]{2008ApJ...677..593W} Walkowicz, L.~M., Johns-Krull, C.~M., \& Hawley, S.~L.\ 2008, \apj, 677, 593 
\bibitem[Woods et al.(2004)]{2004ApJ...605..378W} Woods, P.~M., et al.\ 2004, \apj, 605, 378
\bibitem[Wright et al.(2011)]{2011ApJ...730...93W} Wright, J.~T., et al.\ 2011, \apj, 730, 93
\bibitem[Yung \& Demore(1999)]{1999ppa..conf.....Y} Yung, Y.~L., \& Demore, W.~B.\ 1999, Photochemistry of planetary atmospheres / Yuk L.~Yung, William B.~DeMore.~New York : Oxford University Press, 1999.~ QB603.A85 Y86 1999
\bibitem[Yung et al.(1988)]{1988Icar...73..516Y} Yung, Y.~L., Drew, W.~A., Pinto, J.~P., \& Friedl, R.R.\ 1988, Icarus, 73, 516 
\bibitem[Zahnle et al.(2009)]{2009arXiv0911.0728Z} Zahnle, K., Marley, M.~S., \& Fortney, J.~J.\ 2009, arXiv:0911.0728 
\bibitem[Zahnle et al.(2009)]{2009ApJ...701L..20Z} Zahnle, K., Marley, M.~S., Freedman, R.~S., Lodders, K., \& Fortney, J.~J.\ 2009, \apjl, 701, L20
\bibitem[Zehe et al. (2001)]{2009CAP}Zehe, M.J. and Gordon, S. and McBride, B.J. \ 2001, CAP: a computer code for generating tabular thermodynamic functions from NASA lewis coefficients, National Aeronautics and Space Administration, Glenn Research Center


\end{thebibliography}
\end{document}